\documentclass[twocolumn,showpacs,amsfonts,aps,prc,nofootinbib]{revtex4}

\usepackage{mathrsfs}
\usepackage{epsfig}
\usepackage{amssymb}
\usepackage{amsfonts}
\usepackage{latexsym}
\usepackage{amsthm}

\usepackage{bm}
\usepackage{graphicx}
\usepackage{amsmath}
\usepackage{hyperref}
\usepackage{slashed}
\usepackage[font=small,textfont=normal,format=plain,labelfont=bf,slc=false,justification=centerlast]{caption}
\usepackage{epstopdf}
\usepackage[usenames,dvipsnames]{color}
\usepackage{soul}

\usepackage{subcaption}
\usepackage{cleveref}
\def\la{\langle}
\def\ra{\rangle}

\def\eq{{\,=\,}}

\begin{document}
\setstcolor{red}

%%%%%%%%%%%%%%%%%%%%%%%%Front Matter%%%%%%%%%%%%%%%%%%%%
%%%%%%%%%%%%%%%%%%%%%%%%%%%%%%%%%%%%%%%%%%%%%%%%%%

%%%%%%%%%%%%%%%%%%%%%%%%%%%%
\title{Initial state fluctuations in collisions between light and heavy ions} 
%%%%%%%%%%%%%%%%%%%%%%%%%%%%

\author{Kevin Welsh}
\affiliation{Department of Physics, The Ohio State University,
  Columbus, Ohio 43210-1117, USA}
\author{Jordan Singer}
\affiliation{Department of Physics, The Ohio State University,
  Columbus, Ohio 43210-1117, USA}
%\author{Brian Baker}
%\affiliation{Department of Physics, The Ohio State University,
%  Columbus, Ohio 43210-1117, USA}
\author{Ulrich Heinz}
\email[Correspond to\ ]{heinz.9@osu.edu}
\affiliation{Department of Physics, The Ohio State University,
  Columbus, Ohio 43210-1117, USA}
  
\begin{abstract}
In high energy collisions involving small nuclei (p+p or x+Au collisions where x=p, d, or $^3$He) the fluctuating size, shape and internal gluonic structure of the nucleon is shown to have a strong effect on the initial size and shape of the fireball of new matter created in the collision. A systematic study of the eccentricity coefficients describing this initial fireball state for several semi-realistic models of nucleon substructure and for several practically relevant collision systems involving small nuclei is presented. The key importance of multiplicity fluctuations in such systems is pointed out. Our results show large differences from expectations based on conventional Glauber model simulations of the initial state created in such collisions.
\end{abstract}


\date{\today}

\maketitle

%%%%%%%%%%%%%%%%%%%%%%%%%%%%%%%%%%%%%%%%%%%%%%%%%%

%%%%%%%%%%%%%%%%%%%%%%%%%%%%%%%%%%%%%%%%%%%%%%%%%%
\section{Introduction and overview}
\label{sec1}
%%%%%%%%%%%%%%%%%%%%%%%%%%%%%%%%%%%%%%%%%%%%%%%%%%

Ultra-relativistic heavy-ion collisions at the Relativistic Heavy Ion Collider (RHIC) and the Large Hadron Collider (LHC) have provided strong evidence for fluid dynamic behavior of the hot and dense matter created in the collision \cite{:2003dz}. Recently similar flow-like features have also been observed in collisions between small and large nuclei (p+Au/Pb, d+Au, $^3$He+Au \cite{Adare:2014keg, QM15, Abelev:2012ola, Aad:2012gla, CMS:2012qk, Chatrchyan:2013nka, Werner:2013ipa, Werner:2013tya, Bozek:2014cya, Schenke:2014gaa, Romatschke:2015gxa, Koop:2015trj}), and even in very high multiplicity p+p collisions at the LHC \cite{Khachatryan:2010gv, Bozek:2010pb, Werner:2010ss, Werner:2013tya, Habich:2015rtj, Dusling:2015gta}. While hydrodynamic models have been very successful in achieving a quantitatively accurate description of essentially all soft hadron data (momentum spectra and two-particle correlations of both unidentified charged and identified hadrons with transverse momenta below about 2\,GeV) obtained from the collisions between heavy nuclei (Au+Au, Pb+Pb, Cu+Cu and Cu+Au), a similar convergence between theory and experiment has not yet been achieved in collisions involving small nuclei. In these situations, it appears that uncertainties about the internal structure of the nucleon and, related to that, about the fluctuating initial conditions for the spatial distribution of energy and entropy created in the collision degrade significantly the predictive power of the available dynamical evolution models. 

Event-by-event fluctuations of the initial density distribution of highly excited matter created in collisions between large nuclei reflect mostly the stochastic fluctuations of the positions of the nucleons inside the colliding nuclei at their point of impact \cite{Miller:2003kd,Miller:2007ri}. The resulting density fluctuations have a natural length scale of a nucleon radius. Spatial inhomogeneities and anisotropies of the initial density distribution can be quantified by the complex eccentricity coefficients
\begin{equation}
\label{eq1}
\mathcal{E}_n\equiv\epsilon_n e^{i n\Phi_n} 
= - \frac{\int r dr d\varphi\, r^n e^{i n \varphi} s(r,\varphi)}
            {\int r dr d\varphi \,r^n s(r,\varphi)} \quad(n\geq2)
\end{equation}
associated with the initial entropy density profile in the $(r,\varphi)$ plane perpendicular to the beam direction.\footnote{%
	We model the initial entropy rather than energy deposition. 
	After thermalization, the two are related by the equation of state of the medium. 
	In \cite{Qiu:2011iv} it was shown that the differences between eccentricities defined 
	with entropy and energy density weights are small.}
For collisions between large nuclei, nucleon substructure has no significant effect on these eccentricities even though it renders the density profiles more spiky on sub-nucleonic length scales \cite{Schenke:2012wb, Moreland:2012qw}. Such a substructure exists, however, since the distribution of strongly interacting matter inside a nucleon at the time when the nucleons collide is inhomogeneous, due to quantum fluctuations of the quark and gluon fields that participate in the interaction between the colliding nucleons. These result in spatial inhomogeneities in the transverse plane of the amount of beam energy lost by the colliding nucleons and deposited in the collision zone. We show here that this sub-nucleonic structure and its event-by-event fluctuations have a strong effect on the mean and variance of the initial-state eccentricity coefficients which, when propagated through a hydrodynamic evolution model, affect the means and variances of the final anisotropic flow coefficients $V_n =\{e^{in\phi}\}$ (where $\phi$ denotes the azimuthal angle around the beam direction of emitted particles and $\{\dots\}$ denotes the average over all particles of interest from a single event).      

In this paper we model two sources of nucleon substructure, both of them based on the insight \cite{VanHove:1974wa} that at very high collision energies the production of new matter near midrapidity is dominated by the interaction between low-$x$ gluons (where $x$ is the fraction of the proton light-cone momentum carried by the gluon) in the projectile and target nuclei, and that its spatial distribution in the transverse plane therefore tracks the transverse distribution of the glue in the colliding nuclei at the time of impact. The two specific pictures we explore here are the following:

(i) We can consider the three valence quarks in the nucleon as large-$x$ sources of low-$x$ gluon fields. In this picture, the low-$x$ gluon field clouds surrounding the valence quarks create a spatially inhomogeneous gluon field distribution inside the nucleon, distributed around three valence quark centers \cite{Schlichting:2014ipa} whose positions at the time of impact fluctuate from event to event. This model involves two sub-nucleonic length scales, the width $\sigma_q$ of the distribution in the transverse plane of the valence quark positions and the width $\sigma_g$ of the gluon field lumps carried by each valence quarks. These length scales are constrained by the condition that their squares must add up to the mean squared nucleon radius (see below). In addition to the quantum fluctuations in the spatial positions of the colliding patches of glue, we allow for fluctuations in the amount of energy lost by the colliding nucleons and deposited near midrapidity when their gluons interact. These fluctuations are fit to the measured multiplicity distribution in p+p collisions.
   
(ii) We can try to directly account for gluon field quantum fluctuations inside a nucleon, without tying them explicitly to valence quark sources, or inside the clouds carried around by the valence quarks in model (i), by following the ideas of \cite{Muller:2011bb} based on the Color Glass Condensate picture of low-$x$ gluon fields inside nucleons and nuclei \cite{McLerran:1993ni}. In this approach, the sub-nucleonic gluon field fluctuations are characterized by an amplitude and a single transverse correlation length which are both predicted by the model \cite{Muller:2011bb}. The transverse correlation length depends on the gluon saturation momentum $Q_s$ of the model whose value is controlled by the longitudinally projected gluon density in the transverse plane. $Q_s$ is on average largest in the center of the nucleon or nucleus and falls off towards its edge, so it is a local quantity that depends on the position in the transverse plane. In Ref.~\cite{Moreland:2012qw} a numerical algorithm was developed to modulate locally the smooth average transverse density distribution of the colliding nuclei or nucleons, with a stochastic fluctuation factor whose statistical properties reflect the mean amplitude and transverse correlation length of these gluon field fluctuations. However, that implementation cannot handle a local variation of the transverse correlation length $a(\bm{r})\sim1/Q_s(\bm{r})$;  it works instead with a constant value for $\bar{Q}_s$ that reflects the average saturation momentum of the incoming nuclei. We will use the same approximation here. The corresponding average transverse correlation length $\bar{a}\sim1/\bar{Q}_s$ is of sub-nucleonic size and of the order of 0.3\,fm (0.2\,fm) for top RHIC (LHC) energies.

We focus in this paper on the mean values of magnitudes of the initial ellipticity $\epsilon_2$ and triangularity $\epsilon_3$, which are known to be approximately linearly related to the elliptic and triangular flows, $v_2$ and $v_3$, of the finally emitted hadrons. The linear relation between $\epsilon_n$ and $v_n$ for $n\eq2,3$ was checked numerically in hydrodynamic simulations for collisions between large nuclei and found to hold to good approximation over a wide range of collision centralities, except for very peripheral collisions. We will not perform any hydrodynamic simulations in this work, leaving such studies for a future publication, but will draw some conclusions based on the results for the eccentricity coefficients that assume that this linear relation also holds in collisions involving small projectile and/or target nuclei. This assumption will be tested in upcoming work.

As will be demonstrated below, we find very significant effects of sub-nucleonic fluctuations on the eccentricities of the initially produced matter distribution in collisions involving small nuclei (in particular in p+p collisions). However, the effects arising from quark subdivision of nucleons as described in model (i) and from implementing fluctuating gluon field substructure on protons as described in model (ii) are found to be characteristically different. Quark-subdivision, as implemented in model (i), allows for effective radius fluctuations of the proton, similar to the model studied in \cite{Alvioli:2013vk}, while model (ii) does not. This difference has significant effects on the initial eccentricities. This will be demonstrated in Sec.~\ref{sec2}. Following this demonstration, we will therefore focus in the rest of the paper on the predictions of model (i), taking into account only the fluctuating quark substructure of nucleons, using a smooth Gaussian distribution for the gluon cloud associated with each valence quark (i.e. ignoring additional gluon field fluctuations of these gluon clouds).         

The surprising experimental results from p+Pb collisions at the LHC mentioned above have recently led to intense theoretical interest in proton substructure and subnucleonic density fluctuations. In addition to the works already mentioned, we should refer the reader to the recent revival of  the wounded quark Glauber model \cite{Adler:2013aqf, Loizides:2016djv, Bozek:2016kpf} and the recent study of the effect of proton shape fluctuations on coherent and incoherent diffraction in e+p collisions \cite{Mantysaari:2016ykx}. Quark substructure of nucleons has also recently been suggested as an explanation \cite{Albacete_IS2016} of the so-called ``hollowness'' effect \cite{Arriola:2016bxa} (a depletion of inelastic collision strength in zero impact parameter relative to somewhat off-central p+p collisions at high energies) extracted from an analysis of high-energy p+p collisions at the LHC. We will comment below on similarities and differences with other published work at the appropriate places. 

The rest of this paper is organized as follows: In Sec.~\ref{sec2} we introduce our model implementation of three types of quantum fluctuations in the production of new matter near midrapidity in nuclear collisions: fluctuations of nucleon positions within the colliding nuclei, quark-subdivision of nucleons with fluctuating quark positions, and gluon field fluctuations inside the gluon clouds characterizing a nucleon or its valence quarks. In all three cases we also implement the observed multiplicity fluctuations in nucleon-nucleon collisions, ensuring that the multiplicity distributions measured in p+p collisions are correctly reproduced. We discuss the concept and operational definition of collision centrality in p+p collisions and, for each of the three models, study the characteristic features of the initially produced entropy density profiles as well as the centrality dependence of their ellipticity and triangularity in such collisions.  

%%%%%%%%%%%%%%%%%%%%%%%%%%%%%%%%%%%%%%%%%%%%%%%%%%
%%%%%%%%%%%%%%%%%%%%%%%%%%%%%%%%%%%%%%%%%%%%%%%%%%
\section{Initial state models for collisions involving small nuclei}
\label{sec2}
\vspace*{-2mm}
%%%%%%%%%%%%%%%%%%%%%%%%%%%%%%%%%%%%%%%%%%%%%%%%%%

%%%%%%%%%%%%%%%%%%%%%%%%%%%%%%%%%%%%%%%%%%%%%%%%%%
\subsection{Quantum fluctuations on the nucleon level}
\label{sec2a}
\vspace*{-2mm}
%%%%%%%%%%%%%%%%%%%%%%%%%%%%%%%%%%%%%%%%%%%%%%%%%%
%%%%%%%%%%%%%%%%%%%%%%%%%%%%%%%%%%%%%%%%%%%%%%%%%%
\subsubsection{Fluctuating nucleon positions}
\label{sec2a1}
\vspace*{-2mm}
%%%%%%%%%%%%%%%%%%%%%%%%%%%%%%%%%%%%%%%%%%%%%%%%%%

Ignoring sub-nucleonic degrees of freedom, fluctuations on the nucleon level in the initial density profile arise from two sources: quantum fluctuations of the nucleon positions within a nucleus at the time of impact (discussed in this subsection), and fluctuations in the beam energy fraction lost per colliding nucleon and deposited near midrapity. Beam energy loss and midrapidity energy deposition are processes that occur only when the density distributions of strongly interacting matter carried by projectile and target nucleons overlap in the transverse plane; in this sense, the distribution in the transverse plane of energy density deposited near midrapidity is the result of a position measurement of the nucleons inside the colliding nuclei at the time of nuclear impact. Since the ground state wave function of the colliding nuclei is not an eigenstate of the positions of its constituent nuclei, the result of this position measurement fluctuates stochastically from event to event. These fluctuations give rise to peaks and depressions in the deposited energy density whose positions fluctuate from event to event. The height of the peaks fluctuates additionally due to fluctuations in the ``intensity'' of each nucleon-nucleon collision, arising from quantum fluctuations in the quark and gluon field strengths that make up the strongly interacting matter of each nucleon. These latter fluctuations manifest themselves as multiplicity fluctuations in p+p collisions (i.e. as fluctuations in the multiplicity of newly produced particles created in such collisions).  

We use the Monte Carlo Glauber model to sample, event by event, the distribution of nucleon positions inside the nuclei. For large nuclei, such as Au or Pb, the positions of their constituent nucleons are sampled independently from a Woods-Saxon distribution, imposing a minimum inter-nucleon distance (``hard core diameter'') of 0.9\,fm \cite{Luzum:2013yya,Shen:2014vra} to account for repulsive two-nucleon correlations \cite{Alvioli:2009ab}. The parameters of the Woods-Saxon distribution of the nucleon centers are renormalized \cite{Filip:2007tj,Hirano:2012kj,Shen:2014vra} relative to the measured Woods-Saxon nuclear density distribution, to account for the finite size of the nucleons with an assumed Gaussian nucleon density distribution\footnote{We use the notation $\vec{\bf{r}}$ for 3-dimensional vectors and $\bm{r}$ for 2-dimensional vectors in the transverse plane.}
\begin{equation} 
\label{eq2}
	\rho_{n}(\vec{\bf{r}}) = \frac{e^{-\vec{\bf{r}}^2/(2B)}}{(2\pi B)^{3/2}}.
\end{equation}
The probability of two nucleons (one from the projectile, the other from the target nucleus), whose trajectories are separated in the transverse plane by an impact parameter $\bm{b}$, to suffer an inelastic collision is determined by the nucleon-nucleon overlap function (normalized to unity)
\begin{equation}
\label{eq3}
     T_{nn}(b) = \int d^2r\, T_n(\bm{r})\,T_n(\bm{r}{-}\bm{b}) = \frac{e^{-b^2/(4B)}}{4\pi B},
\end{equation}
where
\begin{equation}
\label{eq4}
     T_{n}(\bm{r}) = \int_{-\infty}^\infty dz\, \rho_n(\bm{r},z) = \frac{e^{-r^2/(2B)}}{2\pi B},
\end{equation}
is the nucleon thickness function in the transverse plane (also normalized to 1), obtained by integrating the nucleon density distribution along the beam direction $z$. For a pair of nucleons with transverse positions $\bm{r}_i$ and $\bm{r}_j$ the collision probability is taken to be \cite{Heinz:2011mh}
\begin{equation}
\label{eq5}
	P_{ij} \equiv P(r_{ij}) = 1-e^{-\sigma_{gg} T_{nn}(|\bm{r}_i{-}\bm{r}_j|)},
\end{equation}
where $\sigma_{gg}\propto B$ is the gluon-gluon cross section, with the proportionality constant fixed such that $P(b)$ is normalized to the total inelastic nucleon cross section $\sigma_{_\mathrm{NN}}^\mathrm{inel}$ \cite{Heinz:2011mh}:
\begin{equation}
\label{eq6}
	\int d^2b \, P(b) = \sigma_{_\mathrm{NN}}^\mathrm{inel}.	
\end{equation}
A good fit to measurements is obtained by setting \cite{Shen:2014vra}
\begin{equation}
\label{eq7}
	B\left(\sqrt{s}\right) = \frac{\sigma_{_\mathrm{NN}}^\mathrm{inel}\left(\sqrt{s}\right)}{8\pi},	
\end{equation}
which implies that the nucleon density (\ref{eq2}) grows with increasing collision energy in proportion to the inelastic nucleon-nucleon cross section. We use $\sqrt{B}\eq0.408$\,fm for RHIC collisions at $\sqrt{s}\eq200$\,GeV and $\sqrt{B}\eq0.516$\,fm for LHC collisions at $\sqrt{s}\eq5020$\,GeV. 

We note that computing the nucleon-nucleon collision probability with a Gaussian nucleon thickness function (\ref{eq4}) differs from the popular prescription (see, for example, Ref.~\cite{Nagle:2013lja}) of using disk-like nucleons that scatter with unit probability when the disks overlap ever so weakly but cannot scatter at all for transverse separations of more than the disk diameter. Our prescription allows for a non-zero chance of two centrally colliding nucleons to pass through each other unfazed and for a small but non-zero probability of inelastic collision even for impact parameters that are significantly larger than the equivalent disk diameter. This matters in very peripheral nuclear collisions, but much more so for the ``centrality'' dependence of p+p and light-on-heavy collisions, as will be discussed below.

For small nuclei such as d and $^3$He, the nucleon positions are sampled differently, to better account for the known nuclear wave functions. For deuterons we first sample the relative distance $r_{np}$ between the proton and neutron from a probability density obtained from the Hulthen wave function 
\begin{align} 
\label{eq8} 
	P(r)_{np}&=4\pi r^2 |\Phi(r_{np})|^2,
\nonumber\\
	\Phi(r_{np})&=\sqrt{\frac{\alpha \beta(\alpha{+}\beta)}{2\pi(\alpha{-}\beta)^2}}
	\frac{e^{-\alpha r_{np}}-e^{-\beta r_{np}}}{r_{np}},
\end{align}
where $\alpha\eq0.228$\,fm$^{-1}$ and $\beta\eq1.18$\,fm$^{-1}$ \cite{Hulthen}. We then construct a 3-dimensional vector $\vec{\bf{r}}_{np}\eq{r}_{np}(\sin\theta\cos\phi,$ $\sin\theta\sin\phi, \cos\theta)$ by assigning it the length $r_{np}$ and an arbitrary direction, using uniformly sampled values for $\cos\theta$ and $\phi$, and put the neutron and proton at positions $\vec{\bf{r}}_{n,p}=\bm{b}\pm\frac{1}{2}\vec{\bf{r}}_{np}$. We finally write the deuteron matter density as a sum of two Gaussians,
\begin{equation}
\label{eq9}
   \rho_d(\vec{\bf{r}}) = \frac{1}{(2\pi B)^{3/2}}
   \left(e^{-|\vec{\bf{r}}{-}\vec{\bf{r}}_p|^2/(2B)} + e^{-|\vec{\bf{r}}{-}\vec{\bf{r}}_n|^2/(2B)}\right),
\end{equation}
which corresponds to the deuteron thickness function
\begin{eqnarray}
\label{eq10}
   T_d(\bm{r}) &=& T_n(|\bm{r}{-}\bm{r}_n|) + T_p(|\bm{r}{-}\bm{r}_p|)  
\nonumber\\   
  &=&   \frac{1}{2\pi B} \sum_{i=n,p} e^{-|\bm{r}{-}\bm{r}_i|^2/(2B)}.
\end{eqnarray}

For $^3$He and t nuclei, we used a set of 14,000 samplings of the positions of the three nucleons in these nuclei generated from realistic 3-body ground state wave functions obtained from Green function Monte Carlo calculations using the AV18 + UIX model interaction \cite{Carlson:1997qn}. This set of 3-nucleon configurations was kindly provided by J. Lynn and J. Carlson \cite{Nagle:2013lja}. We sampled these triplets of nucleon positions randomly, rotated the configuration randomly in 3-space, and shifted each of the three nucleons in the transverse plane by $\bm{b}$. Finally, we placed a Gaussian nucleon density distribution (\ref{eq2}) at each of the three nucleon positions, obtaining the $^3$He thickness function
\begin{eqnarray}
\label{eq11}
   T_{{^3}\mathrm{He}}(\bm{r}) = \frac{1}{2\pi B} \sum_{i=1,2,3} e^{-|\bm{r}{-}\bm{r}_i|^2/(2B)}.
\end{eqnarray}

To compute the midrapidity energy deposition, we go through all pairs $(ij)$ of projectile and target nucleons and mark both nucleons in the pair as ``wounded'' with probability $P_{ij}$ from Eq.~(\ref{eq5}). A wounded nucleon remains wounded even if in a subsequent encounter with a different nucleon it suffers no inelastic collision. Each wounded nucleon at tranverse position $\bm{r}_i$ is taken to deposit entropy near midrapidity with transverse area density profile $\sim \frac{1}{2\pi B}\exp[-|\bm{r}{-}\bm{r}_i|^2/(2B)]$. The total deposited entropy density in the transverse plane is
\begin{equation}
\label{eq12}
   s(\bm{r}) = \frac{\kappa_s}{\tau_0} \sum_{i=1}^{N_w} \frac{e^{- |\bm{r}{-}\bm{r}_i|^2/(2B)}}{2\pi B},
\end{equation}
where $N_w$ is the total number of wounded nucleons from both of the colliding nuclei, $\tau_0$ is the longitudinal proper time at which the deposited entropy materializes, and $\kappa_s$ is a normalization factor to be determined later by requiring the total multiplicity density $dN_\mathrm{ch}/dy$ of finally emitted charged particles, averaged over many central collision events, to agree with experiment. The value of $\kappa$ is irrelevant for most of the analyses presented in the rest of this paper.

%%%%%%%%%%%%%%%%%%%%%%%%%%%%%%%%%%%%%%%%%%%%%%%%%%
%%%%%%%%%%%%%%%%%%%%%%%%%%%%%%%%%%%%%%%%%%%%%%%%%%
\subsubsection{Nucleon-nucleon multiplicity fluctuations}
\label{sec2a2}
\vspace*{-2mm}
%%%%%%%%%%%%%%%%%%%%%%%%%%%%%%%%%%%%%%%%%%%%%%%%%%

An entropy deposition model which assumes that each wounded nucleon deposits the same total entropy in the transverse plane cannot account for the distribution of multiplicities observed in p+p collisions. To correct for this, the entropy profile that each wounded nucleon deposits is allowed to fluctuate by an overall factor $\gamma$ sampled from a Gamma distribution
\begin{equation} 
\label{eq13}
	P_\Gamma(\gamma;k,\theta) = \frac{\gamma^{k-1} e^{-\gamma/\theta}}{\theta^k \Gamma(k)}
\end{equation}
where $k$ and $\theta$ are the so-called shape and scale parameters of the Gamma distribution. With this modification Eq.~(\ref{eq12}) is replaced by
\begin{equation}
\label{eq14}
   s(\bm{r}) = \frac{\kappa_s}{\tau_0} 
   \sum_{i=1}^{N_w} \gamma_i\, \frac{e^{- |\bm{r}{-}\bm{r}_i|^2/(2B)}}{2\pi B} ,
\end{equation}
with a total initial entropy per unit rapidity of
\begin{equation}
\label{eq15}
   \frac{dS}{dy} = {\kappa_s} \sum_{i=1}^{N_w} \gamma_i \,.
\end{equation}
Due to the properties of the Gamma distribution, $dS/dy$ is also Gamma distributed, with shape parameter $k_\mathrm{eff}=N_w k$ and scale parameter $\theta_\mathrm{eff}=\kappa_s\theta$.

The final multiplicity distribution is obtained from the Gamma-distributed initial entropy $dS/dy$ by an additional folding with a Poisson distribution that accounts for finite multiplicity fluctuations in the hadronization process, yielding a negative binomial final multiplicity distribution \cite{Broniowski:2007nz}. For this one needs to know the mean multiplicity $dN_\mathrm{ch}/dy$ at freeze-out, since this number controls the mean and variance of the Poisson distribution. We take $dN_\mathrm{ch}/dy$ from the experimental data when comparing with measured multiplicity distributions.
 
\begin{figure}[!ht] 
	\includegraphics[width = 0.95\linewidth]{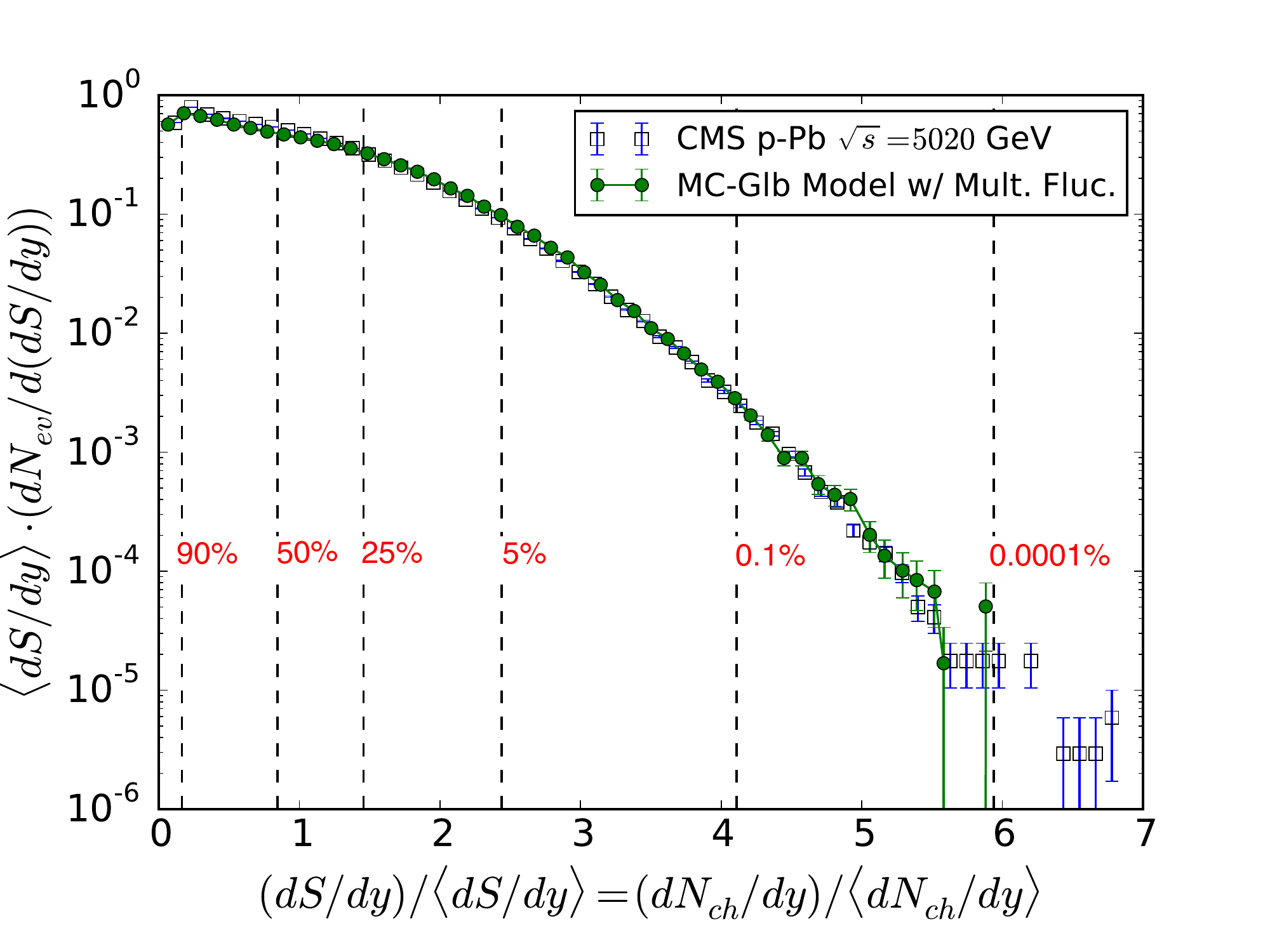}
	\caption{Comparison of the normalized charged multiplicity distribution in p+Pb collisions
	at $\sqrt{s_{_\mathrm{NN}}}=5020$\,GeV measured by CMS \cite{Chatrchyan:2013nka, 
	CMS-HIN-12-015} (blue squares) with the Gaussian nucleon MC-Glauber model including 
	pp multiplicity fluctuations (green circles), as described in the text. 500,000 p+Pb initial 
	entropy density distributions were sampled for the model simulations; to increase statistics, 
	for each event the Poisson distribution was oversampled 5 times, resulting in 2.5 million
	theoretical data points. Vertical dashed lines labelled by red numbers define ``centrality'' 
	classes as fractions of the total number of events.
	\label{F1}
	}
\end{figure}

In Ref.~\cite{Shen:2014vra} it was shown that a good description of the multiplicity distributions measured in p+p collisions can be achieved with $k\eq1/\theta$ and by fixing the scale parameter to $\theta\eq0.61$ for RHIC collisions at $\sqrt{s}\eq200\,A$\,GeV and to $\theta\eq0.75$ for p+p collisions with $\sqrt{s}\eq5020\,A$\,GeV at the LHC. In Fig.~\ref{F1} we show the multiplicity distribution for proton-lead collisions at the same LHC energy resulting from this approach, compared with CMS data. Whereas an MC-Glauber model without pp multiplicity fluctuations fails to describe the high-multiplicity tail of the distribution \cite{Shen:2014vra}, their inclusion yields a very good description of the experimental data up to the highest multiplicities.
 
%
%%%%%%%%%%%%%%%% Fig. 1 %%%%%%%%%%%%%%%%%%%%%%%%%%%%%%%%% 
%%%%%%%%%%%%%%%%%%%%%%%%%%%%%%%%%%%%%%%%%%%%%%%%%%
%

%
%%%%%%%%%%%% Fig. 2 %%%%%%%%%%%%%%%%%%%%%%%%%%%%%%%%%%%
\begin{figure*}
	\includegraphics[width = \textwidth]{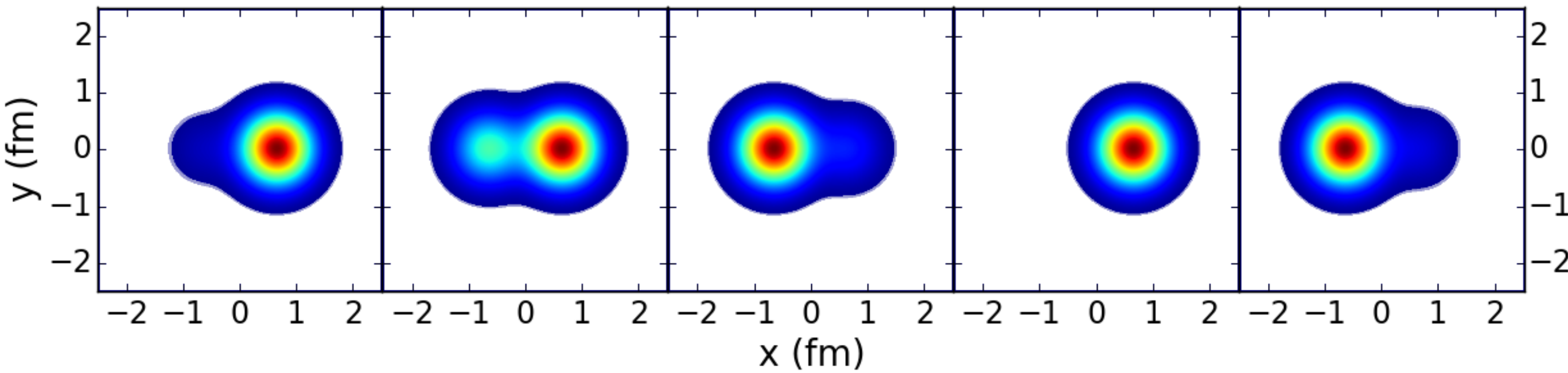}\\
	\caption{Contour plots of the initial entropy density for five randomly selected p+p collisions 
		at $\sqrt{s}\eq200$\,GeV and impact parameter $b\eq1.3$\,fm, computed with the 
		MC-Glauber model using a smooth Gaussian nucleon density profile for collision 
		detection and including multiplicity fluctuations in the deposited entropy.   
		\label{F2}
	}
\end{figure*}

%%%%%%%%%%%%%%%%%%%%%%%%%%%%%%%%%%%%%%%%%%%%%%%%%%
%
Figure~\ref{F2} shows the initial entropy density profiles for 5 randomly selected p+p collisions, colliding at $b\eq1.3$\,fm with $\sqrt{s}\eq200$\,GeV, computed with the model described in this subsection. Without multiplicity fluctuations, the profiles would be mirror symmetric with respect to the reaction plane at $y\eq0$, and all odd eccentricity harmonics (in particular the triangularity $\epsilon_3$) would vanish exactly. In Fig.~\ref{F2} one sees with the naked eye that multiplicity fluctuations induce initial density distributions with generically large triangular deformations in p+p collisions.  

%%%%%%%%%%%%%%%%%%%%%%%%%%%%%%%%%%%%%%%%%%%%%%%%%%
\subsection{Sub-nucleonic quantum fluctuations}
\label{sec2b}
\vspace*{-2mm}
%%%%%%%%%%%%%%%%%%%%%%%%%%%%%%%%%%%%%%%%%%%%%%%%%%

In this subsection we describe how we generalize the MC-Glauber model to account for the quark substructure of nucleons (Sec. \ref{sec2b1}) and for sub-nucleonic gluon field fluctuations (Sec. \ref{sec2b2}).

%%%%%%%%%%%%%%%%%%%%%%%%%%%%%%%%%%%%%%%%%%%%%%%%%%
\subsubsection{Quark subdivision of nucleons}
\label{sec2b1}
\vspace*{-2mm}
%%%%%%%%%%%%%%%%%%%%%%%%%%%%%%%%%%%%%%%%%%%%%%%%%%

Schenke and Schlichting \cite{Schlichting:2014ipa} showed that, when one solves the JIMWLK equation \cite{Balitsky:1995ub,JalilianMarian:1997jx,Mueller:2001uk} for the distribution of gluons at small $x$ generated by three valence quarks at large $x$, the low-$x$ gluons appear in three lumps whose centers track the positions of the valence quarks. The intrinsic length scale of the gluon clouds  (i.e. the ``gluonic radius of a quark'') increases with $\ln(1/x)$, i.e. the gluon clouds become more fuzzy as the rapidity gap between the valence quarks near beam rapidity and the gluons near midrapidity increases.  

We here propose the following model implementation of this picture. We assume that at large $x$ the proton is made of three valence quarks whose positions $\vec{\bf{r}}_i$ ($i\eq1,2,3$) are distributed with a Gaussian probability distribution of width $\sigma_q$ around the nucleon center $\vec{\bf{r}}_n$:
\begin{equation}
\label{eq16}
  P(\vec{\bf{r}}_i{-}\vec{\bf{r}}_n) = 
  \frac{e^{-(\vec{\bf{r}}_i{-}\vec{\bf{r}}_n)^2/(2\sigma_q^2)}}{(2\pi\sigma_q^2)^{3/2}}.
\end{equation}
Each quark $i$ carries with it a Gaussian density distribution $g(\vec{\bf{r}}{-}\vec{\bf{r}}_i)$ of width $\sigma_g$ of low-$x$ gluons that is normalized to 1/3:
\begin{equation}
\label{eq17}
  g(\vec{\bf{r}}{-}\vec{\bf{r}}_i\bigr) = \frac{1}{3}\,
  \frac{e^{-(\vec{\bf{r}}{-}\vec{\bf{r}}_i)^2/(2\sigma_g^2)}}{(2\pi\sigma_q^2)^{3/2}}.
\end{equation}
It is natural to assume that $\sigma_q$ is controlled by confinement and is a fixed fraction of the proton radius in its rest frame, whereas $\sigma_g$ grows with collision energy, leading to the observed growth of the total inelastic nucleon-nucleon cross section as described by Eqs.~(\ref{eq2}), (\ref{eq7}). The two width parameters are constrained by the requirement that, on average, the total glue distribution of the nucleon is given by Eq.~(\ref{eq2}) (which is normalized to 1):
\begin{widetext}
\begin{eqnarray}
\label{eq18}
  \bigl\la\rho_n(\vec{\bf{r}}{-}\vec{\bf{r}}_n)\bigr\ra &=&
  \frac{\sum_{i=1}^3 \int d^3r_1\,P(\vec{\bf{r}}_1{-}\vec{\bf{r}}_n) 
                                \int d^3r_2\,P(\vec{\bf{r}}_2{-}\vec{\bf{r}}_n) 
                                \int d^3r_3\,P(\vec{\bf{r}}_3{-}\vec{\bf{r}}_n) \,
                                \delta(\vec{\bf{r}}_1{+}\vec{\bf{r}}_2{+}\vec{\bf{r}}_3{-}3\vec{\bf{r}}_n)\,
                                g(\vec{\bf{r}}{-}\vec{\bf{r}}_i)}
        {\int d^3r_1\,P(\vec{\bf{r}}_1{-}\vec{\bf{r}}_n) 
                                \int d^3r_2\,P(\vec{\bf{r}}_2{-}\vec{\bf{r}}_n) 
                                \int d^3r_3\,P(\vec{\bf{r}}_3{-}\vec{\bf{r}}_n) \,
                                \delta(\vec{\bf{r}}_1{+}\vec{\bf{r}}_2{+}\vec{\bf{r}}_3{-}3\vec{\bf{r}}_n)}
\nonumber\\                               
  &=&                                                                 
   \frac{e^{-(\vec{\bf{r}}{-}\vec{\bf{r}}_n)^2/(2B)}}{(2\pi B)^{3/2}}.
\end{eqnarray}
\end{widetext} 
Here the $\delta$-function in numerator and denominator ensures that the center of mass of the three valence quarks agrees with $\vec{\bf{r}}_n$, the center of the nucleon. Eq.~(\ref{eq18}) yields the constraint
\begin{equation}
\label{eq19}
   \sigma_g^2 + \frac{2}{3}\sigma_q^2 = B.
\end{equation}
The factor $2/3$ multiplying $\sigma_q^2$ reflects the fact that only for two of the three quarks the positions can be sampled randomly from the probability distribution (\ref{eq16}) whereas the third quark's position is then fixed by the nucleon center of mass. $B$ is fixed experimentally by the collision energy, and we will use Eq.~(\ref{eq19}) to eliminate $\sigma_q=\sqrt{\frac{3}{2}(B{-}\sigma_g^2)}$ and to express all results as a function of the width $\sigma_g$ of the gluon lumps carried around by the quarks.

Introducing the shifted and scaled quark positions $\vec{\bm{\xi}}_i{\,\equiv\,}(\vec{\bf{r}}_i{-}\vec{\bf{r}}_n)/\sigma_q$ and using the $\delta$-function to integrate over the position of the third quark, we can rewrite
Eq.~(\ref{eq18}) as follows:
\begin{widetext}
\begin{equation}
\label{eq20}
  \bigl\la\rho_n(\vec{\bf{r}}{-}\vec{\bf{r}}_n)\bigr\ra =
  \frac{\int d^3\xi_1\,d^3\xi_2\, 
          e^{-\frac{1}{2}\left(\xi_1^2+\xi_2^2 + (\vec{\bm{\xi}}_1+\vec{\bm{\xi}}_2)^2\right)}
          \left[g(\vec{\bf{r}}{-}\vec{\bf{r}}_n{-}\sigma_q\vec{\bf{\xi}}_1) +  
                      g(\vec{\bf{r}}{-}\vec{\bf{r}}_n{-}\sigma_q\vec{\bf{\xi}}_2) +  
                      g\left(\vec{\bf{r}}{-}\vec{\bf{r}}_n{+}\sigma_q(\vec{\bf{\xi}}_1{+}\vec{\bf{\xi}}_2)\right)\right]}
        {\int d^3\xi_1\,d^3\xi_2\,
         e^{-\frac{1}{2}\left(\xi_1^2+\xi_2^2+(\vec{\bm{\xi}}_1+\vec{\bm{\xi}}_2)^2\right)}}.
\end{equation}
\end{widetext} 
This identifies the weight
\begin{equation}
\label{eq21}
  P(\vec{\bm{\xi}}_1,\vec{\bm{\xi}}_2)=
  \frac{e^{-\frac{1}{2}\left(\xi_1^2+\xi_2^2+(\vec{\bm{\xi}}_1+\vec{\bm{\xi}}_2)^2\right)}}
         {\int d^3\xi_1\,d^3\xi_2\,
         e^{-\frac{1}{2}\left(\xi_1^2+\xi_2^2+(\vec{\bm{\xi}}_1+\vec{\bm{\xi}}_2)^2\right)}}
\end{equation}
as the probability distribution to be sampled for the vectors $\vec{\bm{\xi}}_{1,2}$ that determine the positions 
\begin{equation}
\label{eq22}
    \vec{\bf{r}}_{1,2}=\vec{\bf{r}}_n+\sigma_q\vec{\bm{\xi}}_{1,2}, \quad
    \vec{\bf{r}}_3=\vec{\bf{r}}_n-\sigma_q(\vec{\bm{\xi}}_1{+}\vec{\bm{\xi}}_2)
\end{equation}
of the three valence quarks. Each sampled pair $\vec{\bm{\xi}}_{1,2}$ yields a nucleon centered at $\vec{\bf{r}}_n$ with density distribution
\begin{equation}
\label{eq23}
\!\!\!
  \rho_n(\vec{\bf{r}};\vec{\bf{r}}_1,\vec{\bf{r}}_2,\vec{\bf{r}}_3)
  = \sum_{i=1}^3 g(\vec{\bf{r}}{-}\vec{\bf{r}}_i)
  = \sum_{i=1}^3 \frac{e^{-(\vec{\bf{r}}{-}\vec{\bf{r}}_i)^2/(2\sigma_g^2)}}
                                  {(2\pi\sigma_g^2)^{3/2}},
\!\!                                  
\end{equation}
with $\vec{\bf{r}}_i$ from (\ref{eq22}). The corresponding nucleon thickness function (\ref{eq4}) is
\begin{equation}
\label{eq24}
  T_n(\bm{r};\bm{r}_1,\bm{r}_2,\bm{r}_3)
   = \sum_{i=1}^3 \frac{e^{-(\bm{r}{-}\bm{r}_i)^2/(2\sigma_g^2)}}
                                  {2\pi\sigma_g^2},
\end{equation}
where all 3-dimensional vectors $\vec{\bf{r}}$ in (\ref{eq23}) have been replaced by 2-dimensional vectors $\bm{r}$ in the transverse plane. The nucleon-nucleon overlap function (\ref{eq3}), which is needed for evaluating with Eq.~(\ref{eq5}) the collision probability $P_{ij}$ between two nucleons $i$ and $j$ separated by $\bm{b}\eq\bm{r}_i{-}\bm{r}_j$, is given by
\begin{equation}
\label{eq25}
  T_{nn}(\bm{b})= \sum_{k,l=1}^3 \frac{e^{-(\bm{r}_k^{(i)}{-}\bm{r}_l^{(j)}{-}\bm{b})^2/(4\sigma_g^2)}}
                                                            {4\pi\sigma_g^2}.
\end{equation}
Here $\bm{r}_k^{(i)}$ ($k\eq1,2,3$) are the positions of the three valence quarks in nucleon $i$, and similarly for nucleon $j$.  

%
%%%%%%%%%%%%%%%% Fig. 3 %%%%%%%%%%%%%%%%%%%%%%%%%%%%%%%%% 
\begin{figure}[b] 
	\includegraphics[width = \linewidth]{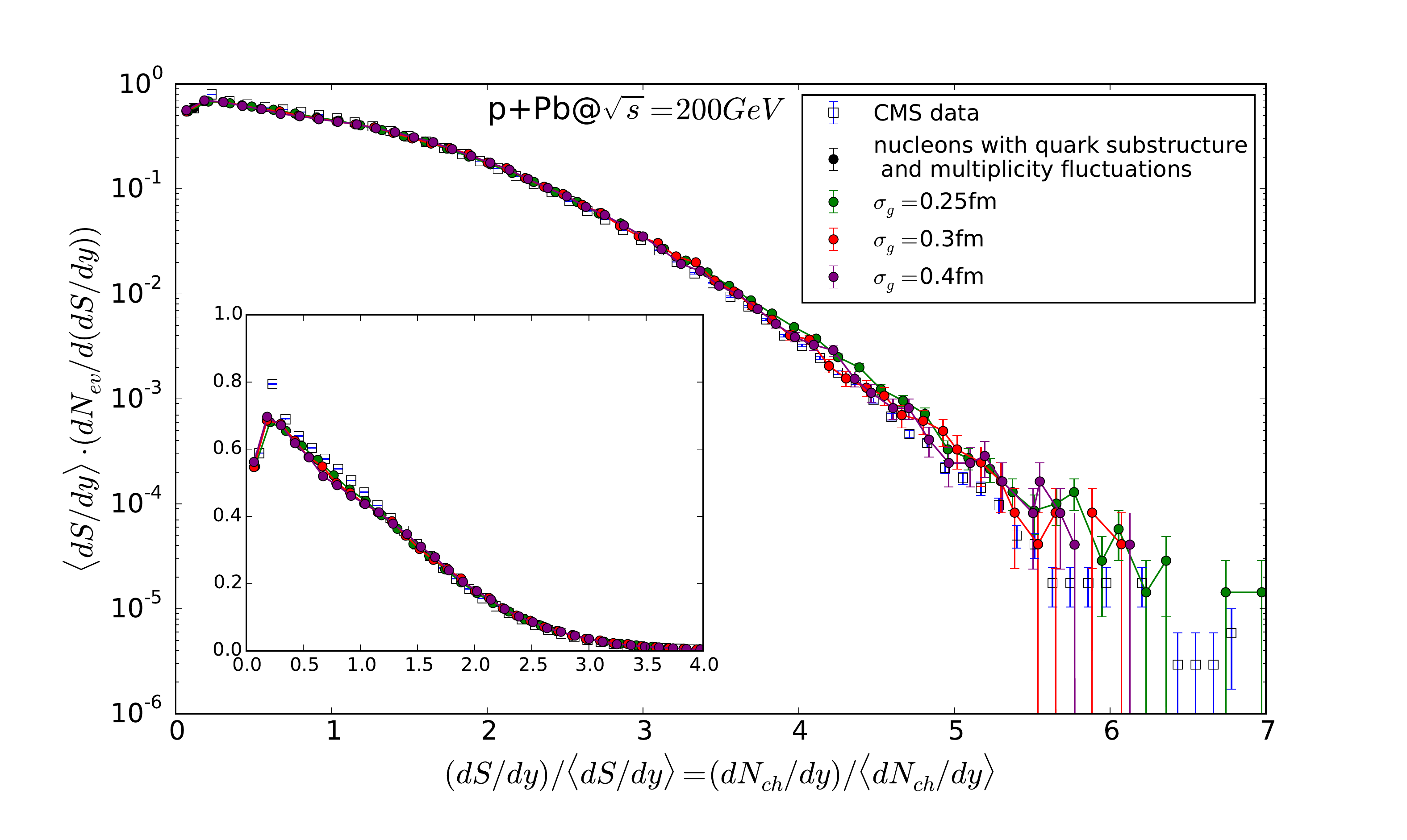}\\
	\caption{Same as Fig.~\ref{F1}, but now the experimental data from CMS are compared with
		simulations in which the protons are subdivided into gluon clouds of width $\sigma_g$,
		with three choices for $\sigma_g$ as given in the legend, located around three valence 
		quarks with fluctuating positions as described in the text. The inset is a linear plot to better 
		see the sensitivity to $\sigma_g$ at low multiplicities.
	\label{F3}
	} 
\end{figure}
%%%%%%%%%%%%%%%%%%%%%%%%%%%%%%%%%%%%%%%%%%%%%%%%%%
%
%
%%%%%%%%%%%% Fig. 4 %%%%%%%%%%%%%%%%%%%%%%%%%%%%%%%%%%%
\begin{figure*}
	\includegraphics[width = \textwidth]{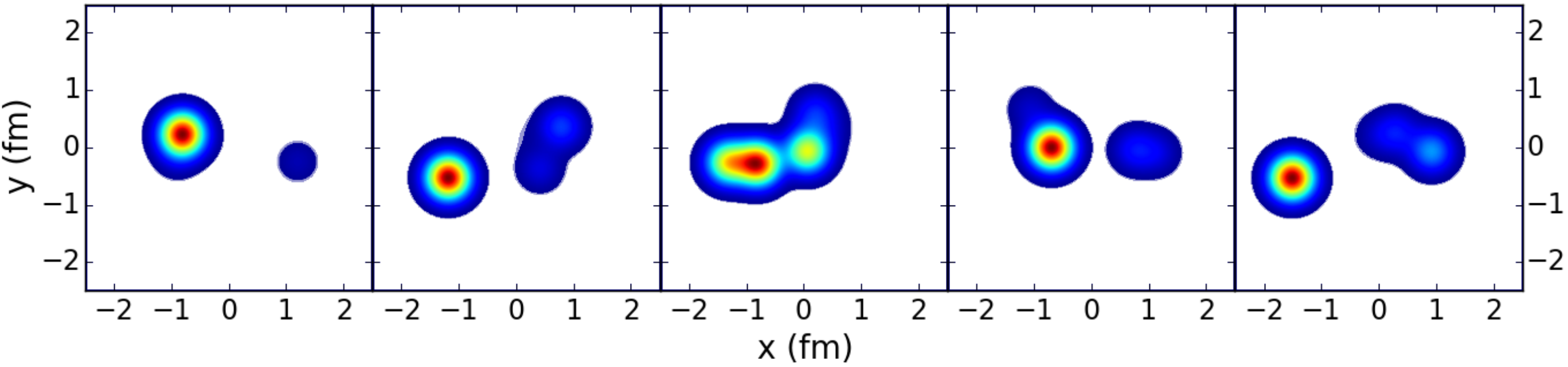}\\
	\caption{Contour plots of the initial entropy density for five randomly selected p+p collisions 
		at $\sqrt{s}\eq200$\,GeV and impact parameter $b\eq1.3$\,fm, computed with the 
		MC-Glauber model using quark subdivision of the nucleon density profile for both 
		collision detection and entropy deposition, including additional multiplicity fluctuations in the 
		deposited entropy. See text for model description and discussion.  
		\label{F4}
	}
\end{figure*}

%%%%%%%%%%%%%%%%%%%%%%%%%%%%%%%%%%%%%%%%%%%%%%%%%%%
%
As before, for each projectile-target nucleon pair $(ij)$ in the two colliding nuclei both nucleons are labelled as ``wounded'' with probability $P_{ij}$. Once a nucleon is wounded, its quark substructure gets broken up, creating rapidly separating color charges connected by strong color fields. This process which involves all three valence quarks, converts a fraction of the nucleon's initial beam energy and deposits it near midrapidity. This amount fluctuates from event to event, and we assume it to fluctuate independently for each valence quark. The deposited entropy density is thus modeled as
\begin{equation}
\label{eq26}
   s(\bm{r}) = \frac{\kappa_s}{\tau_0} 
   \sum_{i=1}^{N_w} \sum_{k=1}^3 \gamma_k^{(i)}\, 
   \frac{e^{-(\bm{r}{-}\bm{r}_k^{(i)})^2/(2\sigma_g^2)}}{2\pi\sigma_g^2},
\end{equation}
where $\gamma_k^{(i)}$ is Gamma distributed with
\begin{equation} 
\label{eq27}
	P_\Gamma(\gamma;k_q,\theta_q) = 
	\frac{\gamma^{k_q-1} e^{-\gamma/\theta_q}}{\theta_q^{k_q} \Gamma(k_q)},
\end{equation}
using the same scale and a thrice smaller shape parameter compared to nucleons without substructure, $\theta_q\eq\theta_n$ and $k_q\eq{k}_n/3$, in order to ensure the same multiplicity distribution in p+p collisions as for the Glauber model without quark substructure. When averaged over many samplings of the valence quark positions and $\gamma$-factors, the mean entropy density deposited per wounded nucleon at position $\bm{r}_n$ is proportional to the Gaussian thickness function (\ref{eq4}) of that nucleon without subdivision:
\begin{eqnarray}
\label{eq28}
   \bigl\la s(\bm{r}) \bigr\ra &=& \left\la\frac{\kappa_s}{\tau_0} 
   \sum_{k=1}^3 \gamma_k^{(n)}\, 
   \frac{e^{-(\bm{r}{-}\bm{r}_k^{(n)})^2/(2\sigma_g^2)}}{2\pi\sigma_g^2}\right\ra
\nonumber\\
   &=& 
   \frac{\kappa_s}{\tau_0} \frac{e^{-(\bm{r}{-}\bm{r}_n)^2/(2B)}}{2\pi B}
\end{eqnarray}

While quark substructure has no effect on the multiplicity distribution in p+p collisions (by construction of our model), it does slightly modify the multiplicity distribution in p+Pb collisions which, as seen in Fig.~\ref{F3}, shows some sensitivity to the choice of the width $\sigma_g$ of the gluon clouds around the valence quarks (and thereby to the variance $\sigma_q^2$ of the fluctuating quark positions). At average multiplicities, small $\sigma_g$ values yield the best description of the experimental data whereas the extreme tail of the measured multiplicity distribution is best described by larger $\sigma_g$ values, even as large as $\sqrt{B}$ (in which case the three valence quarks sit on top of each other and the proton density distribution degenerates to a smooth profile described by a single Gaussian). We choose from here on $\sigma_g\eq0.3$\,fm as a compromise value that yields the best overall description of the measured p+Pb multiplicity distribution at the LHC. It is possible that experimental data on diffraction in $e$+p collisions can be used to more tightly constrain $\sigma_g$, following the methods proposed in \cite{Mantysaari:2016ykx}, but we leave this to a future study.

Figure~\ref{F4} shows the effect of quark subdivision of nucleons on the initial entropy density profiles in p+p collisions at $b\eq1.3$\,fm and {$\sqrt{s}=200$\,GeV. Compared to the case of smooth Gaussian protons shown in Fig.~\ref{F2}, accounting in the collision detection probability for bumpy gluon density distributions within each nucleon appears to favor somewhat more compact distributions of produced entropy density: Since the collision probability increases when the valence quarks from projectile and target overlap, collisions between quark-diquark-like nucleon configurations in which the diquarks face each other, thereby ensuring stronger overlap, are favored.%
\footnote{% 
       We checked and found that computing the collision probability from the quark-subdivided thickness
       function (\ref{eq24}) tends to favor generically collisions with more compact wounded nucleon-nucleon 
       configurations than computing it with the Gaussian thickness function (\ref{eq4}), resulting in 
       5-7\% smaller ellipticities and triangularities in pp collisions. Note that multiplicity fluctuations 
       are only included in the entropy deposition process, not in the calculation of the collision probability.
}
Clearly, the density profiles shown in Fig.~\ref{F4} have a more bumpy character than those without quark subdivision, explaining the larger eccentricity values for collisions between nucleons with quark substructure that we discuss next.

%
%%%%%%%%%%%%%%%%% Fig. 5 %%%%%%%%%%%%%%%%%%%%%%%%%%%%%%%
\begin{figure*}
	\includegraphics[width=0.48\linewidth,height=6.57cm]{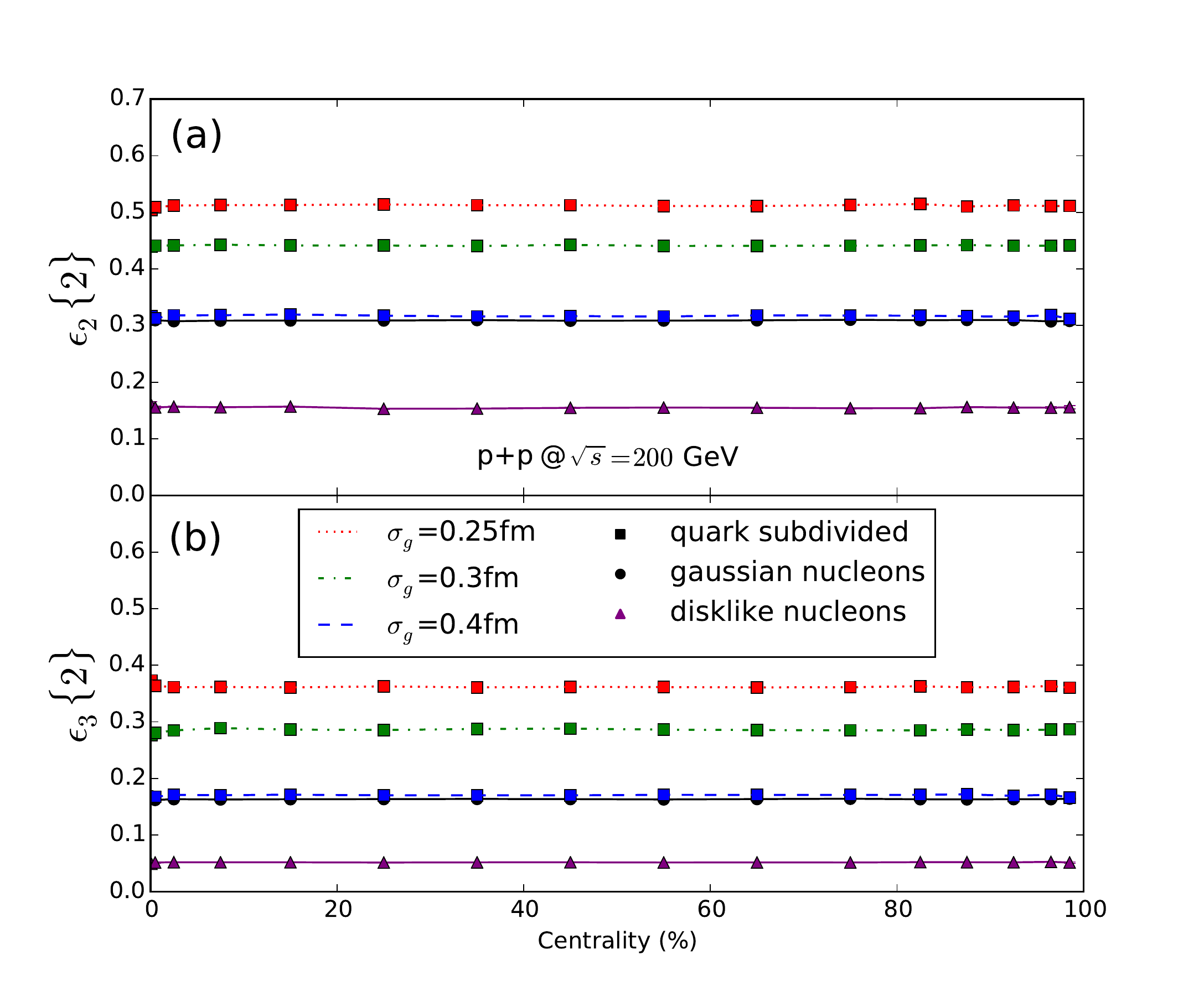}
	\includegraphics[width=0.48\linewidth]{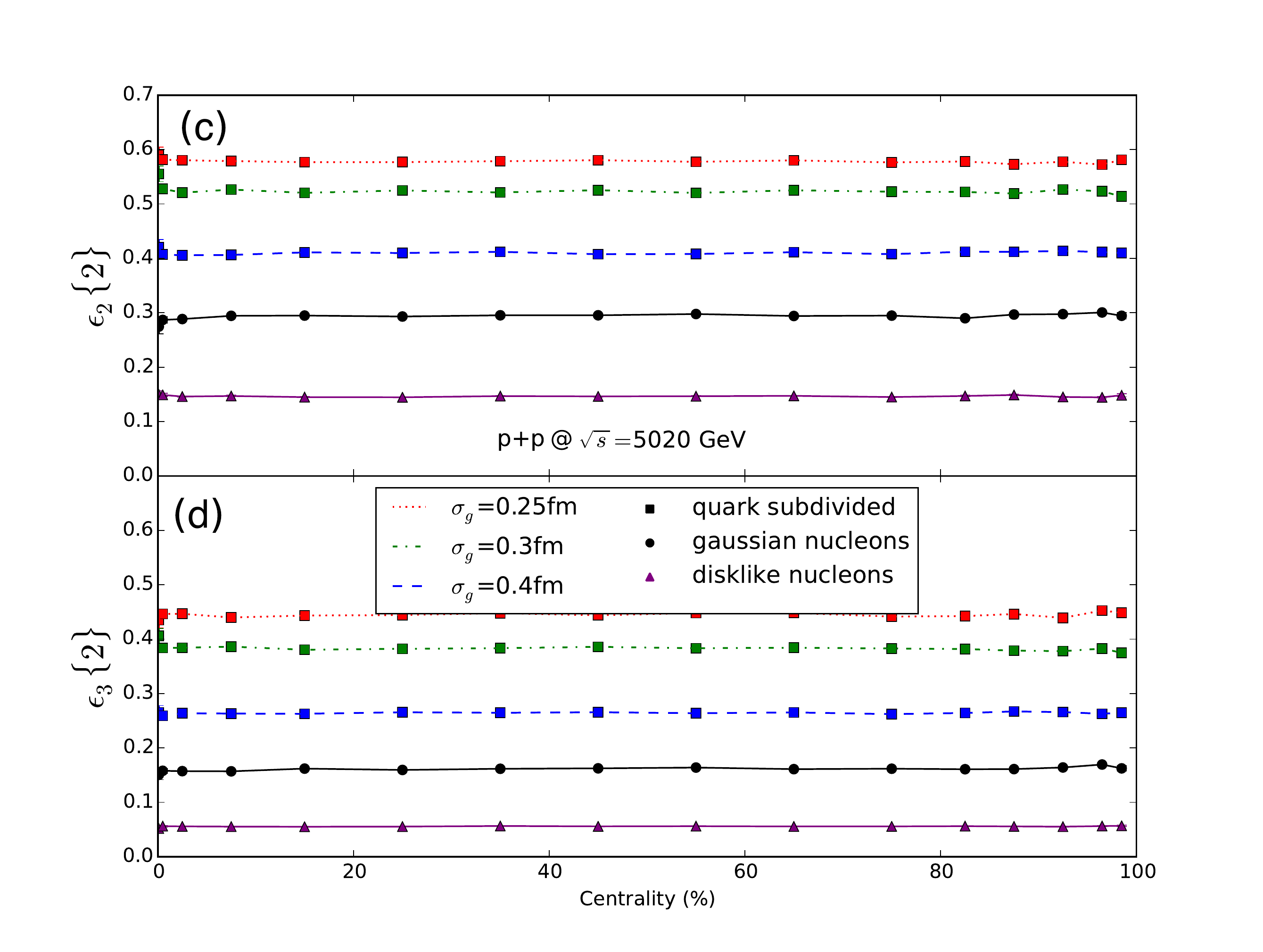}
	\caption{Elliptic and triangular rms eccentricities, $\epsilon_2\{2\}$ (a,c) and $\epsilon_3\{2\}$ (b,d),
		      for p+p collisions at $\sqrt{s}\eq200$\,GeV (a,b) and 5020\,GeV (c,d). See text for 
		      detailed discussion.
         \label{F5}
         }
\end{figure*}
%%%%%%%%%%%%%%%%%%%%%%%%%%%%%%%%%%%%%%%%%%%%%%%%%%%
%

In Figure~\ref{F5} we show the elliptic and triangular eccentricities as a function of collision ``centrality'' (as measured by the total initial entropy $dS/dy$, see Fig.~\ref{F1}), for p+p collisions at $\sqrt{s}\eq200$\,GeV (where $\sqrt{B}\eq0.408$\,fm) in panels (a,b), and for  $\sqrt{s}\eq5020$\,GeV (where $\sqrt{B}\eq0.516$\,fm) in panels (c,d). Let us explain the model versions for which results are shown in the figure. All five model variants include p+p multiplicity fluctuations -- without those, $\epsilon_3$ would be zero by reflection symmetry for nucleons without quark subdivision. The lines labeled ``disk'' (solid purple triangles) use a disk-like collision criterium where, for the purpose of collision detection, the nucleons are represented by black disks whose radius is fixed by the total inelastic nucleon-nucleon cross section, assuming 100\% collision probability when the disks overlap and zero collision probability when they don't. Nucleons that have been wounded by suffering a collision then deposit entropy with a single Gaussian profile as described in Sec.~\ref{sec2a2}. Fig.~\ref{F5} shows that this model generates the smallest ellipticities and triangularities. The black lines with solid black circles uses a collision detection criterium based on Eq.~(\ref{eq5}) with a Gaussian nucleon profile, again followed by entropy deposition as described in Sec.~\ref{sec2a2}. This produces twice larger elliptic and three times larger triangular eccentricities than obtained for disk-like collision detection. This clearly demonstrates the importance of collisions with impact parameters exceeding the disk diameter where only the tails of the Gaussian nucleon distributions overlap. Though rare, these collisions have a strong effect on the eccentricities, boosted by multiplicity fluctuations in the final state. Note that the purple triangles and black circles don't move much by going from RHIC to LHC energies, in spite of the $~20\%$ increase of the proton radius.

%
%%%%%%%%%%%%%%% Fig. 6 %%%%%%%%%%%%%%%%%%%%%%%%%%%%
\begin{figure}[t]
	\includegraphics[width = \linewidth]{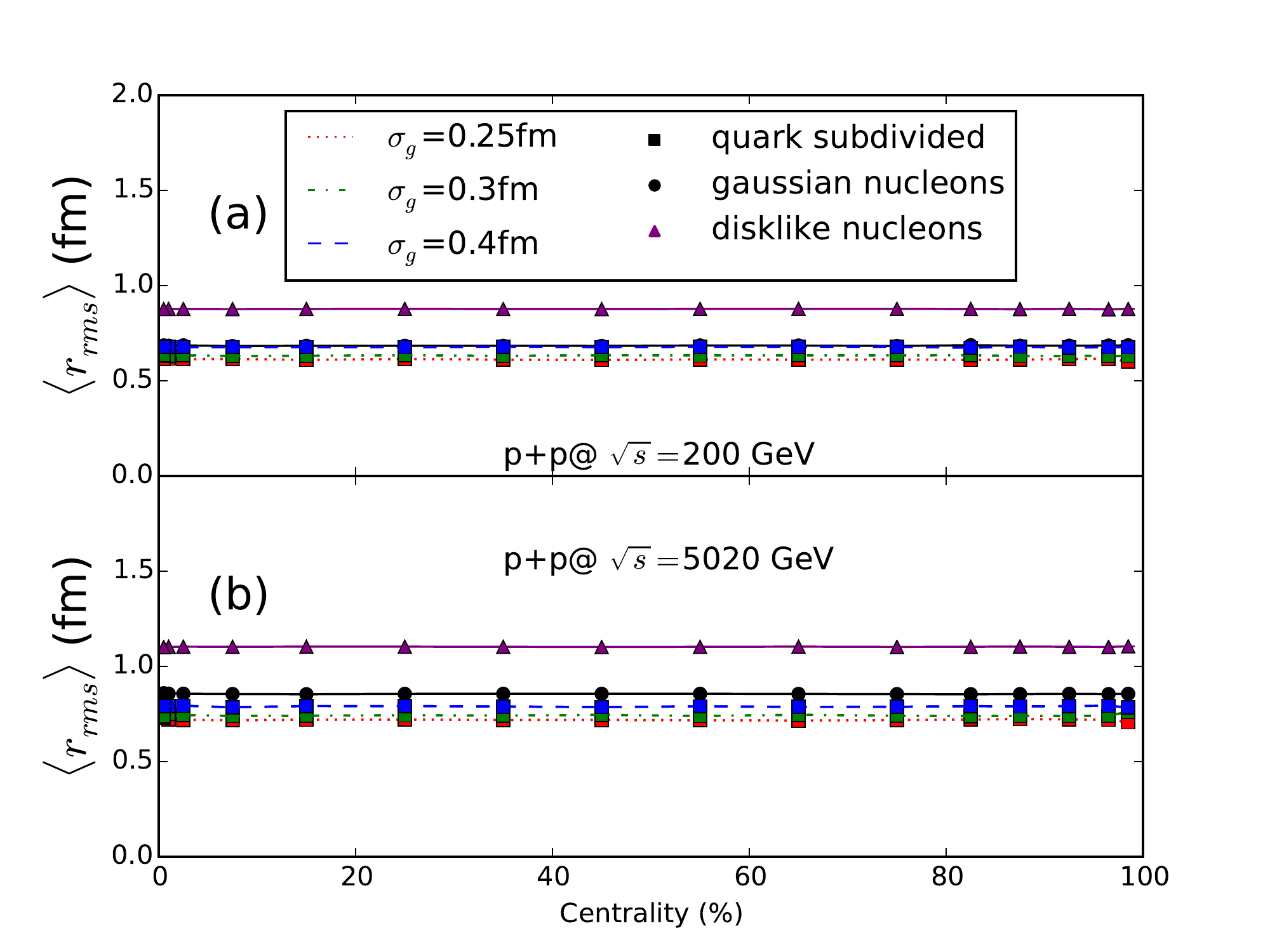}\\
	\caption{Ensemble-averaged rms radii for p+p collisions at $\sqrt{s}\eq200$\,GeV (a) and 
		5020\,GeV (b), as a function of collision centrality, using the same symbols and line styles
		as in Fig.~\ref{F5}. See text for discussion.
		\label{F6}
	}
\end{figure}%
%%%%%%%%%%%%%%%%%%%%%%%%%%%%%%%%%%%%%%%%%%%%%%
%

The remaining three curves in Fig.~\ref{F5} (labelled by red, green, and blue squares connected by dotted, dash-dotted and dashed lines, respectively) show the initial-state eccentricities in collisions between nucleons with valence quark substructure, as described in Sec.~\ref{sec2b1}, for gluon cloud width parameters $\sigma_g\eq0.25$, 0.3, and 0.4\,fm, respectively. The largest of these $\sigma_g$ values almost exhausts the width $\sqrt{B}\eq0.408$\,fm of the Gaussian describing the average nucleon density distribution at RHIC energies, explaining why in this case the eccentricities almost agree with the results for smooth Gaussian nucleons without quark subdivision (see Fig.!\ref{F5}a,b). As $\sigma_g$ decreases, and thus the variance $\sigma_q^2$ of the fluctuating quark positions in the nucleon increases, the fluctuating nucleon densities become more inhomogeneous, resulting in larger eccentricities. The largest eccentricity values would be obtained for pointlike valence quarks ($\sigma_g=0$) with quark position variance $\sigma_q^2\eq\frac{3}{2}B$ (see Eq.~(\ref{eq19})). Note that quark substructure increases $\epsilon_2$ on average by up to 80\% and can more than double the average triangularity $\epsilon_3$. Due to the increasing proton radius $\sqrt{B}$ from RHIC to LHC energies, valence quarks with the same gluon cloud radius $\sigma_g$ experience more freedom at LHC energies for moving around inside the proton, leading to larger elliptic and triangular eccentricities at LHC than at RHIC (c.f. panels (a,b) vs. (c,d) in Fig.~\ref{F5}). 

Figure~\ref{F6} shows the average rms radii of the initial entropy density distributions for the same set of model assumptions studied in Fig.~\ref{F5}, for p+p collisions at RHIC and LHC energies. The traditional disk-like collision detection criterium leads to the largest source radii. Using a Gaussian nucleon profile for computing the collision probability favors more compact collision configurations, resulting in more compact deposited entropy profiles. Further quark-subdivision leads to still smaller initial source radii, with the smallest valance quark gluon clouds (i.e. the biggest variance for the quark positions inside the nucleon) leading to the most compact initial fireball configurations, smaller by about 40\% compared to those calculated with disk-like collision detection. Naturally, such more compact initial density distributions feature larger density gradients, resulting in larger radial flow after hydrodynamic evolution and larger mean transverse momenta of the finally emitted hadrons.  

%
%%%%%%%%%%%%%%% Fig. 7 %%%%%%%%%%%%%%%%%%%%%%%%%%%%
\begin{figure}[t]
	\includegraphics[width = \linewidth]{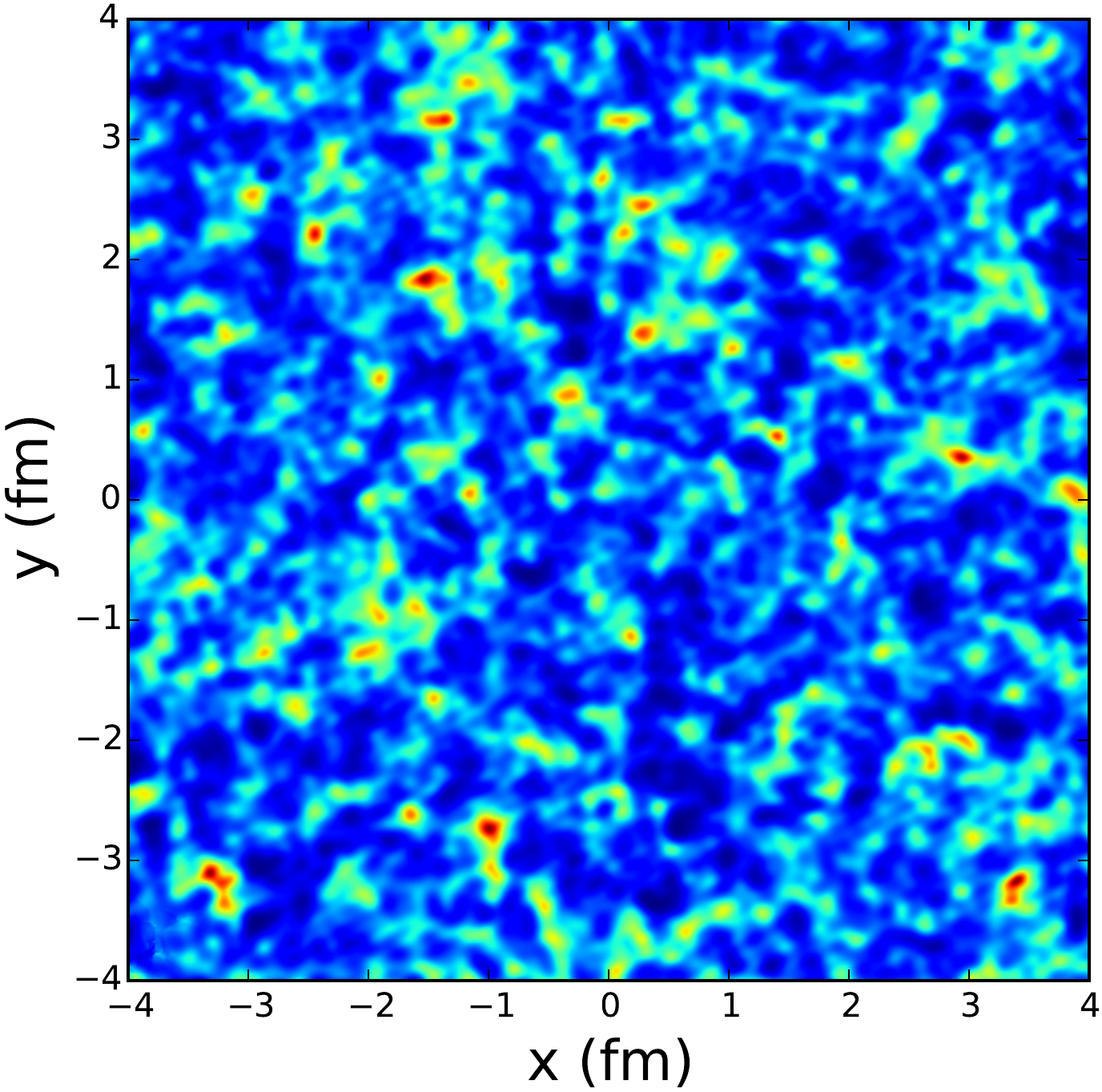}\\
	\caption{A representive sample of the Gamma-distributed random field (with transverse correlation 
		length $\bar a=0.29$\,fm) used for modulating the gluon field strength inside a Gaussian
		nucleon, or within the Gaussian gluon cloud carried along by a quark when nucleons are
		subdivided into valence quarks. 
		\label{F7}
	}
\end{figure}%
%%%%%%%%%%%%%%%%%%%%%%%%%%%%%%%%%%%%%%%%%%%%%%
%

It is worth emphasizing that the method introduced in this subsection to account for quark substructure differs from what has become known as the ``constituent quark Monte-Carlo Glauber model'' \cite{Adler:2013aqf, Loizides:2016djv, Bozek:2016kpf}, by enforcing (through the center-of-mass constraint in Eq.~(\ref{eq18})) the clustering of three quarks each into a nucleon, and by insisting that {\em all three quarks in a wounded nucleon} contribute to the production of new matter at midrapidity. Also, by sampling the valence quark positions from a Gaussian distribution of width $\sigma_q$, we implicitly allow for fluctuations of the nucleon r.m.s. radius. In earlier versions of the Monte Carlo Glauber model such nucleon size fluctuations were introduced via a fluctuating nucleon-nucleon cross section \cite{Alvioli:2013vk}. Both of these features have a significant influence on the initial eccentricities of the fireballs formed in collisions involving nuclei made of just a single or a few nucleons.

%%%%%%%%%%%%%%%%%%%%%%%%%%%%%%%%%%%%%%%%%%%%%%%%%%
\subsubsection{Sub-nucleonic gluon field fluctuations}
\label{sec2b2}
%%%%%%%%%%%%%%%%%%%%%%%%%%%%%%%%%%%%%%%%%%%%%%%%%%

In this subsection explore the consequences of accounting for gluon field fluctuations inside the colliding nucleons for the initial entropy density profile. Following Ref.~\cite{Moreland:2012qw}, we first generate a 2-dimensional field of Gamma distributed random variables with unit mean and transverse correlation length $\bar{a}$.  A representative sample of this field is shown in Fig.~\ref{F7}.
%
%%%%%%%%%%%%%%%%% Fig. 8 %%%%%%%%%%%%%%%%%%%%%%%%%%%%%%%
\begin{figure}[bh!]
	\includegraphics[width = \linewidth]{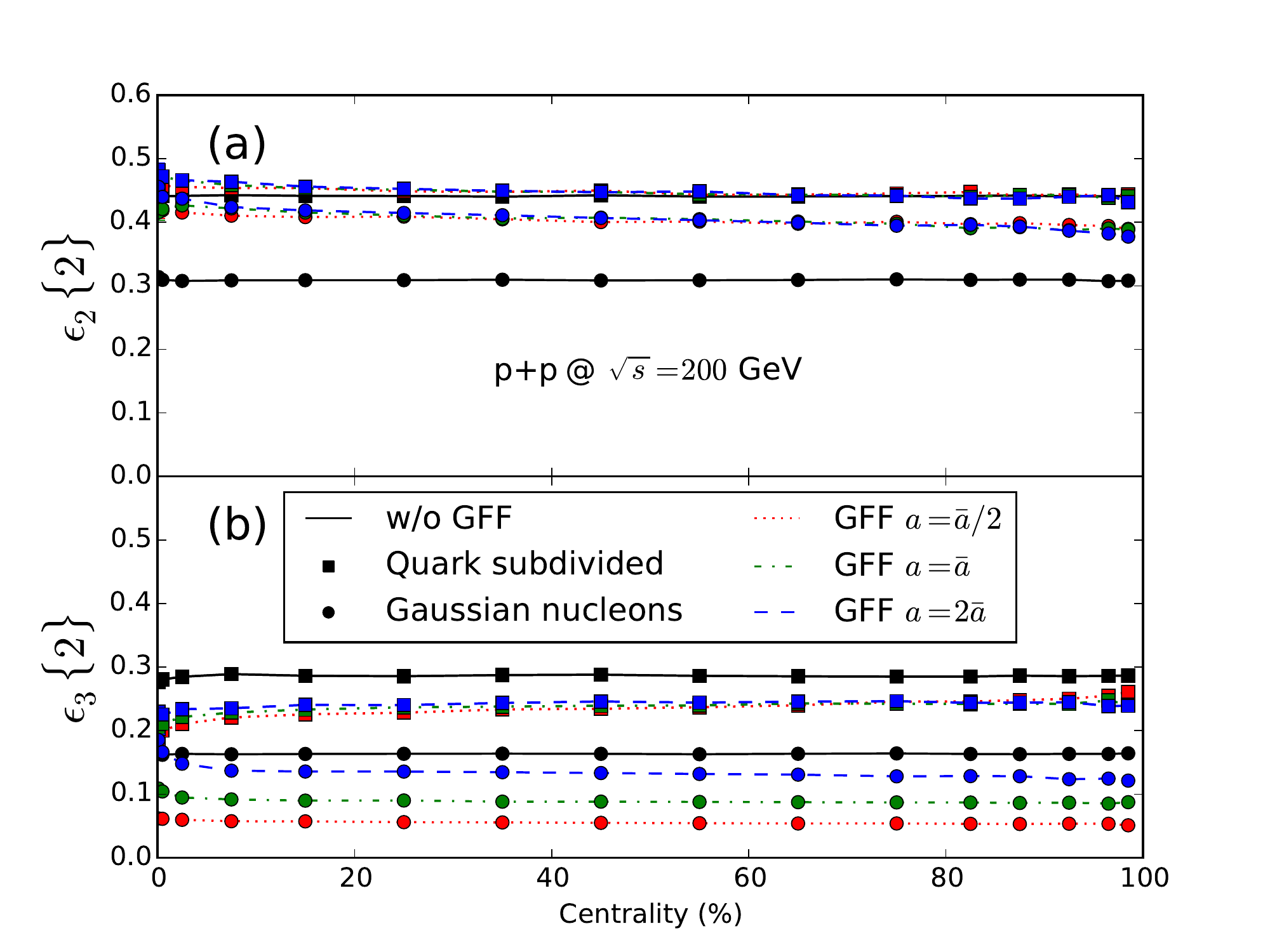}
	\caption{Similar to Fig.~\ref{F5}, but including gluon field fluctuations. The width of the
		gluon cloud carried by valence quarks for quark-subdivided nucleons has been set
		to $\sigma_g\eq0.3$\,fm, and $\bar{a}\eq0.29$\,fm is taken for the transverse gluon 
		field correlation length in collisions at $\sqrt{s}\eq200$\,GeV. See text for detailed 
		discussion.
         \label{F8}
         }
\end{figure}%
%%%%%%%%%%%%%%%%%%%%%%%%%%%%%%%%%%%%%%%%%%%%%%%%%%%
%
%
%%%%%%%%%%%%%%%%%%%% Fig. 9 %%%%%%%%%%%%%%%%%%%%%%%
\begin{figure*}
	\includegraphics[width = 0.85\textwidth]{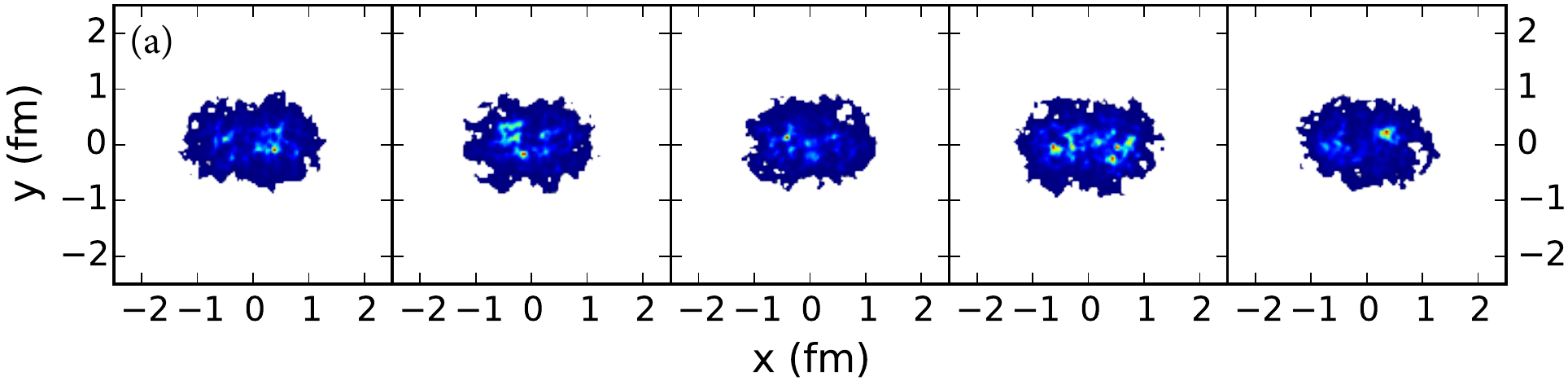}\\[-4ex]
	\includegraphics[width = 0.85\textwidth]{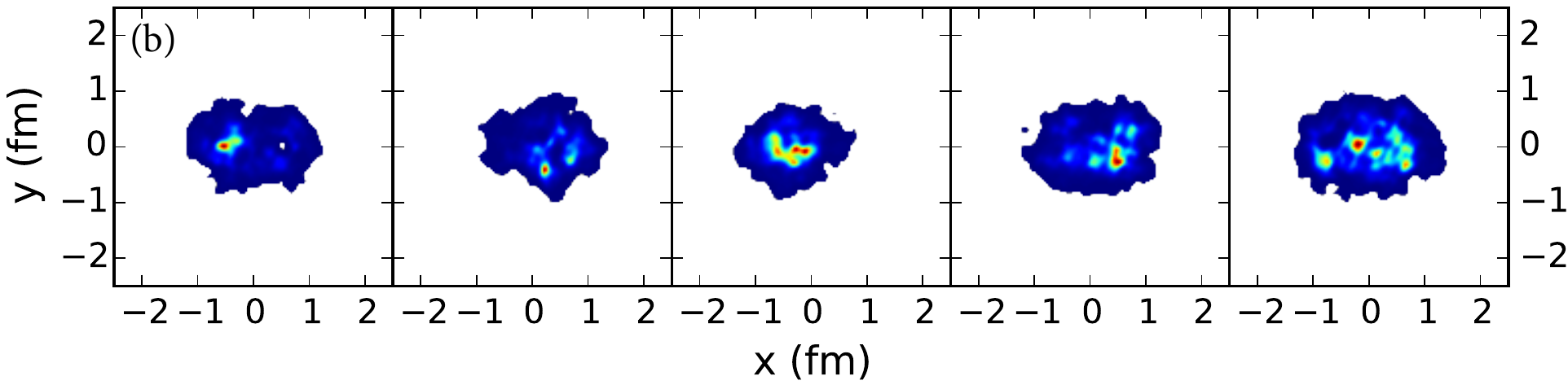}\\[-4ex]
	\includegraphics[width = 0.85\textwidth]{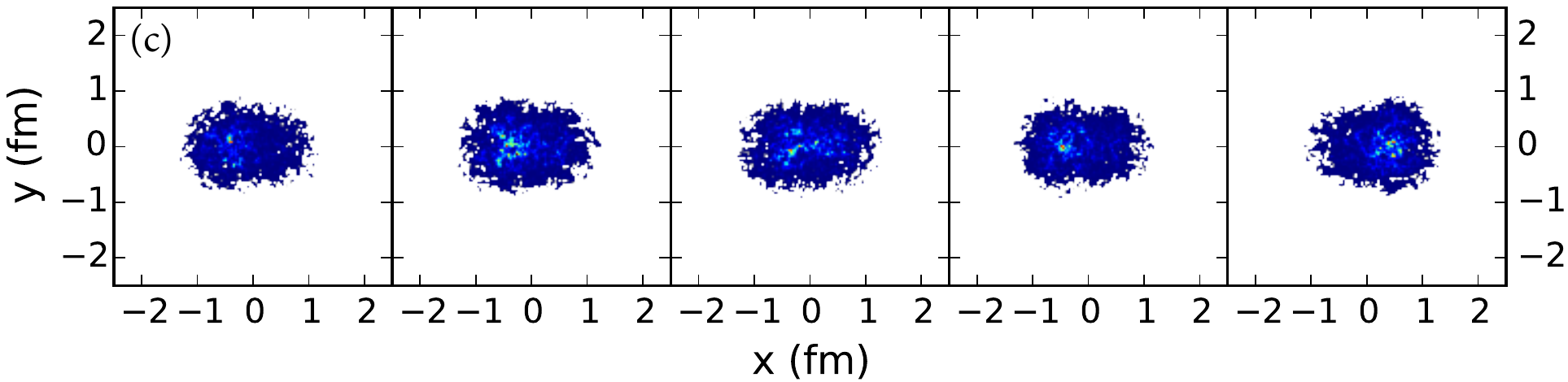}\\[-4ex]
	\includegraphics[width = 0.85\textwidth]{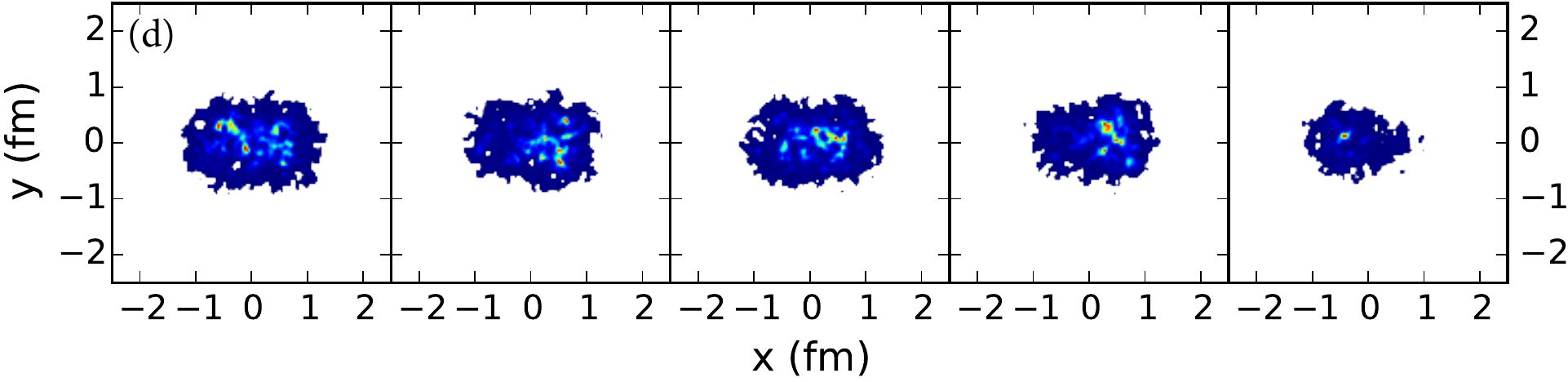}\\[-4ex]
	\includegraphics[width = 0.85\textwidth]{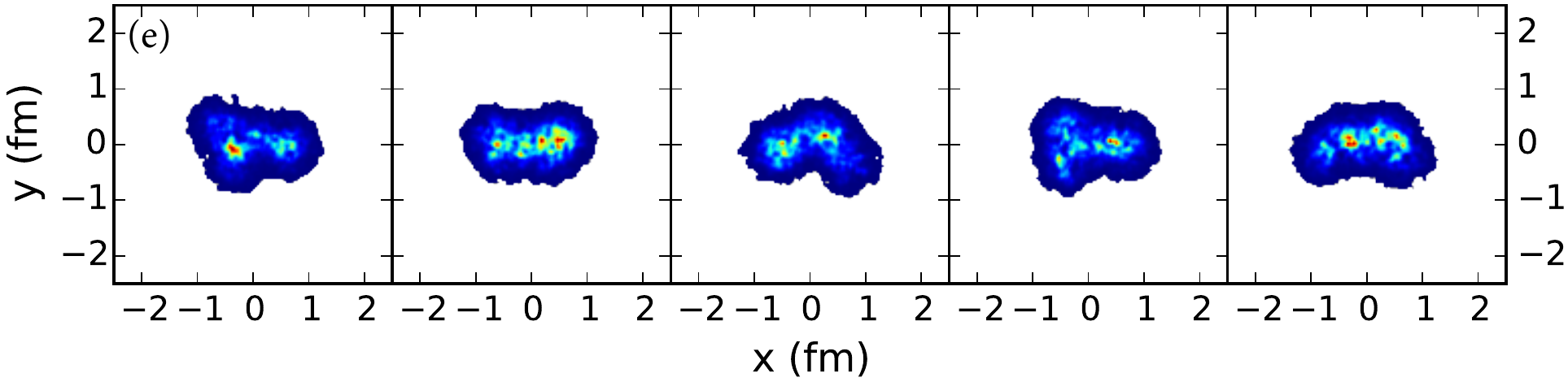}\\[-4ex]
	\includegraphics[width = 0.85\textwidth]{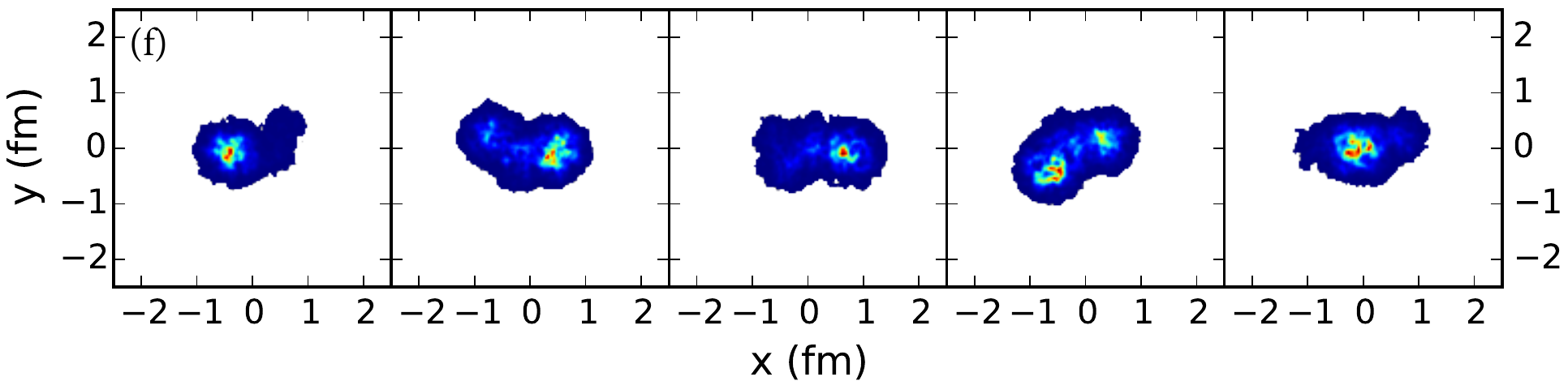}\\
	\caption{Contour plots of the initial entropy density for five randomly selected p+p collisions 
		at $\sqrt{s}\eq200$\,GeV and impact parameter $b\eq1.3$\,fm, computed with the 
		MC-Glauber model with Gaussian or quark-subdivided nucleons whose Gaussian gluon 
		density distributions have been modulated by the Gamma-distributed random field shown
		in Fig.~\ref{F7}. The fluctuated gluon density distributions are used for both collision 
		detection and entropy deposition. In the entropy deposition step multiplicity fluctuations
		are either included or not, as specified in the following: (a) Gaussian nucleons with gluon 
		field fluctuations (GFF) of transverse correlation length $a=\bar a=0.29$\,fm, without multiplicity
		fluctuations (MF); (b) same as (a) except for $a=2\bar a$; (c) same as (a) except for  
		$a=\bar a/2$; (d) same as (a) but including fluctuations in the multiplicity generated by 
		each nucleon; (e) quark-subdivided nucleons with valence quark gluon clouds of width 
		$\sigma_g=0.3$\,fm modulated by GFF with $a=\bar a$ without MF; (f) same as (e) but 
		including independent fluctuations of the multiplicities created by each quark.  
		\label{F9}
	}
\end{figure*}%
%%%%%%%%%%%%%%%%%%%%%%%%%%%%%%%%%%%%%%%%%%%%%%%%
%
As stated in the introduction, we use $\bar{a}\eq0.29$\,fm at $\sqrt{s_{_\mathrm{NN}}}\eq200$\,GeV and $\bar{a}\eq0.2$\,fm at $\sqrt{s_{_\mathrm{NN}}}\eq5020$\,GeV. Let us denote the value of this field at transverse position $\bm{r}$ by $\Gamma(\bm{r})$. To apply the fluctuations, a large (30\,fm$\,\times\,$30\,fm) field was generated from which to sample. For each nucleon $i$ centered at $\bm{r}_i$, a random square section of the field $\Gamma$, with center $\bm{R}_i$ and width $10\sqrt{B}$, is sampled from the grid. The entropy density deposited by each wounded nucleon is then modified by multiplication with the random field as follows:
\begin{equation}
  \label{eq29}
  s(\bm{r}) = s_0(\bm{r})\,\Gamma(\bm{r}{+}\bm{R}_i)\, \Theta\bigl(5\sqrt{B} - |\bm{R}_i-\bm{r}|\bigr),
\end{equation}
where $s_0$ is the original deposited entropy density profile from Eq.~(\ref{eq14}) or (\ref{eq27}) for $N_w\eq1$). 

The effect of sub-nucleonic gluon field fluctuations on the initial elliptic and triangular eccentricities are shown in Fig.~\ref{F8}. How they affect the initially deposited entropy density profiles can be seen in Fig.~\ref{F9} for $b\eq1.3$\,fm pp collisions at $\sqrt{s}\eq200$\,GeV. As described in the caption, all combinations (gluon field fluctuations imprinted in pp collisions for Gaussian and quark-subdivided nucleons with and without multiplicity fluctuations and for three choices of the transverse correlation length of the fluctuating gluon fields) have been studied.  We point out the qualitatively different characteristics of the entropy density profiles generated by quark-subdivided nucleons (panels (e) and (f) in Fig.~\ref{F9}; see also Fig.~\ref{F4} for the case without gluon field fluctuations) compared to those generated by Gaussian nucleons with (panels (a-d) of Fig.~\ref{F9}) and without gluon field fluctuations. The radial fluctuations of the nucleon size in the quark-subdivision model allow for stronger spatial deformations, generating larger elliptic and triangular eccentricities, which are impossible to mimic by starting with a Gaussian nucleon profile of fixed radius and modulating it by gluon field fluctuations.   

Figure \ref{F8} shows the effect of sub-nucleonic gluon field fluctuations with and without quark-subdivision on the elliptic and triangular eccentricities of the initial entropy density profiles generated in pp collisions at RHIC energies. Only the realistic case which includes fluctuations in the multiplicities generated by the nucleons or their valence quarks is considered. One sees that the addition of gluon field fluctuations does not affect the almost perfect centrality independence of the two eccentricities in pp collisions. This centrality (i.e. multiplicity) independence is due to the large variance of the multiplicity fluctuations in pp collisions. We will show further below that these large multiplicity fluctuations completely invalidate any geometric interpretation of the measured centrality variable (i.e. event multiplicity) in terms of impact parameter or geometric overlap between the colliding objects in any collision where at least one of the two colliding nuclei is small (i.e. contains less than a handful of nucleons).  

Whereas quark subdivision increases the initial eccentricities $\epsilon_{2,3}$ by significant factors, as discussed in Fig.~\ref{F5} above (c.f. circles and squares connected by solid lines in Fig.~\ref{F8}), the addition of gluon field fluctuations (dashed, dotted and dash-dotted lines for different choices of the transverse gluon field correlation length as specified in the legend) affects the ellipticity and triangularity differently. For Gaussian nucleons (circles in Fig.~\ref{F8}), gluon field fluctuations {\em increase} the ellipticity $\epsilon_2$ by about 30\%, roughly independent of the transverse gluon field correlation length $a$ (Fig.~\ref{F8}a) while they {\em decrease} the triangularity $\epsilon_3$ (Fig.~\ref{F8}b) by an amount that depends sensitively on the parameter $a$. For quark-subdivided nucleons (squares in Fig.~\ref{F8}), gluon field fluctuations have practically no effect on the ellipticity while decreasing the triangularity by about 15-20\%, independent of $a$. 

Clearly, the strongest effect on the eccentricities is generated by the quark subdivision process, with its concomitant radial fluctuations of the entropy density fluctuations. Gluon field fluctuations lead to an additional increase of the uncertainty range for the triangularities of the initial entropy density distributions in pp collisions. Mindful of the fact that our prescription for including gluon field fluctuation effects is incomplete since it does not properly account for the multiscale nature of these fluctuations, caused by the local variation of the saturation momentum across the transverse plane, we set gluon field fluctuations aside for the rest of the paper, focussing on quark subdivision as the driver for sub-nucleonic density fluctuations in the initial state of nuclear collisions.

%
%%%%%%%%%%%%%%%%% Fig. 10 %%%%%%%%%%%%%%%%%%%%%%%%%%%%%%%
\begin{figure*}
	\includegraphics[width = 0.45\linewidth]{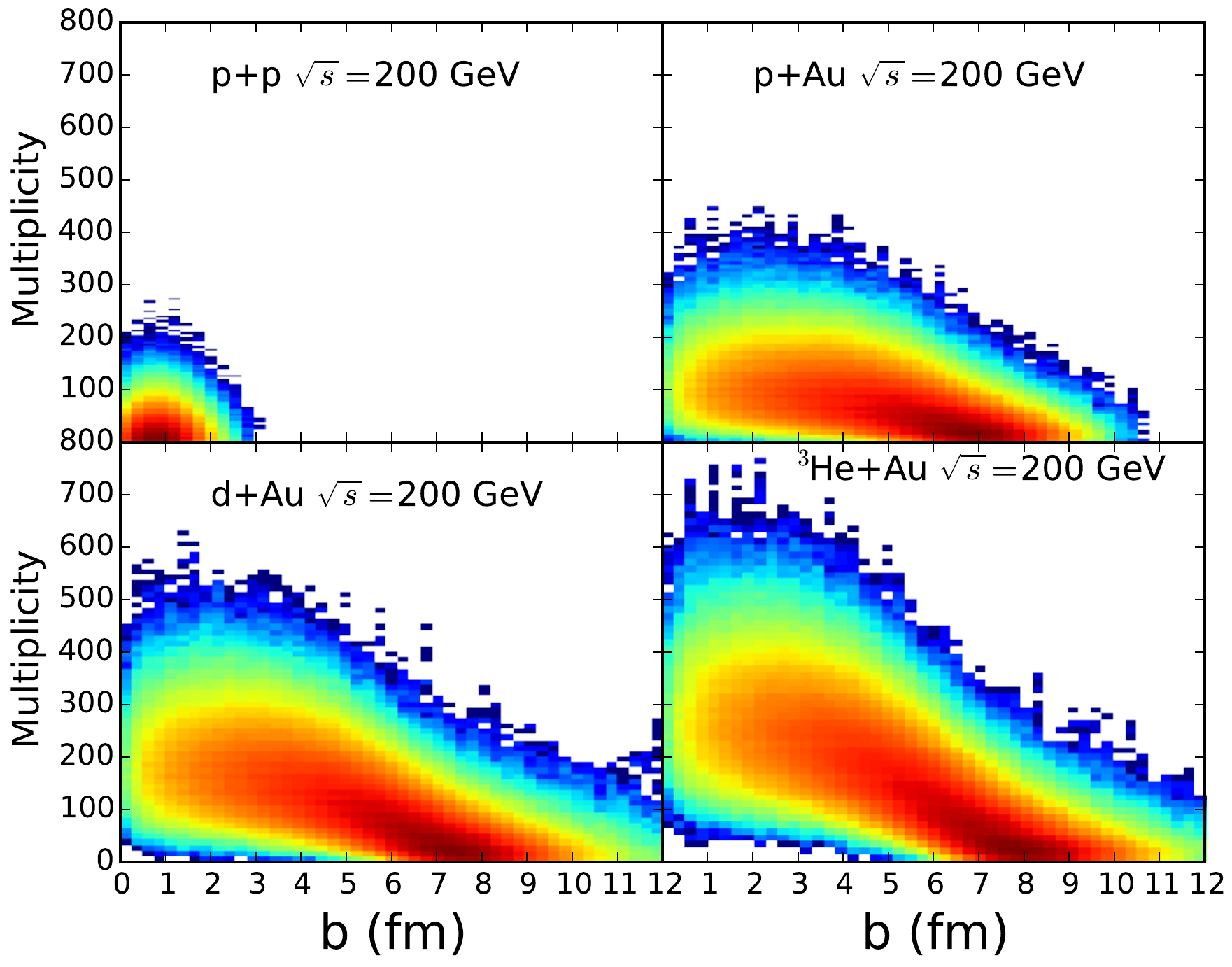}
	\includegraphics[width = 0.45\linewidth]{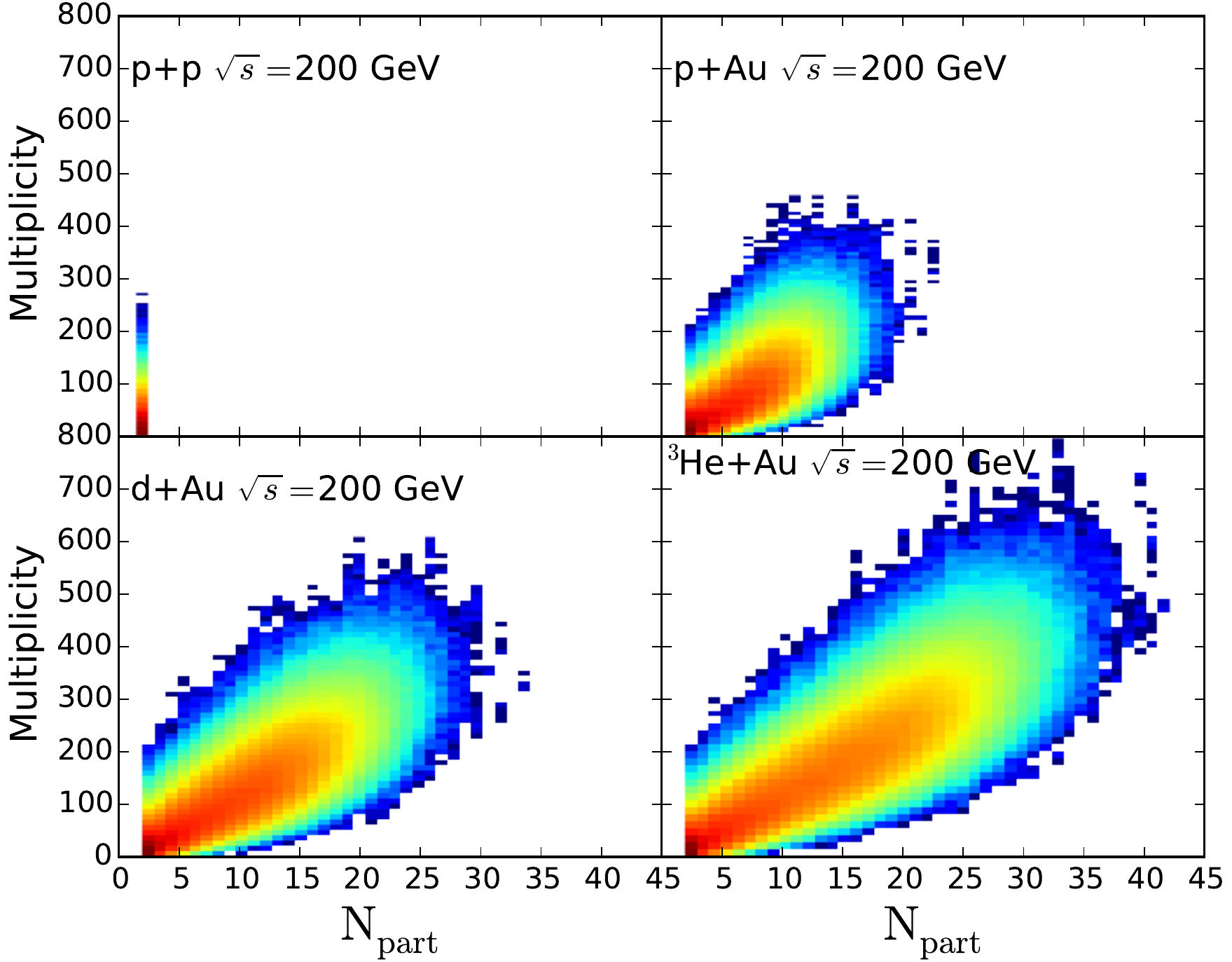}
	\caption{False-color scatter plot of multiplicity $dS/dy{\,\propto\,}dN_\mathrm{ch}/dy$ vs. impact 
	parameter $b$ (a) and vs. the number of participant nucleons $N_\mathrm{part}$ (b) in p+p, p+Au, 
	d+Au and $^3$He+Au collisions at $\sqrt{s}\eq200\,A$\,GeV. Quark-subdivided nucleons with 
	width $\sigma_g\eq0.3$\,fm for the valence quark gluon clouds were used to produce this figure.  
         \label{F10}
         }
\end{figure*}%
%%%%%%%%%%%%%%%%%%%%%%%%%%%%%%%%%%%%%%%%%%%%%%%%%%%
%

%%%%%%%%%%%%%%%%%%%%%%%%%%%%%%%%%%%%%%%%%%%%%%%%%%
%%%%%%%%%%%%%%%%%%%%%%%%%%%%%%%%%%%%%%%%%%%%%%%%%%
\section{Results for $\mathrm{p{+}p}$ and $\mathrm{x{+}Au}$ collisions}
\label{sec3}
%%%%%%%%%%%%%%%%%%%%%%%%%%%%%%%%%%%%%%%%%%%%%%%%%%

For each of the initial state options discussed in Sec.~\ref{sec2}, we generated between 500,000 and 1,000,000 initial entropy density profiles for each collision type, x+Au at $\sqrt{s}\eq200$\,GeV (per nucleon pair) and x+Pb at $\sqrt{s}\eq5.02$\,TeV (per nucleon pair) where x $\in$ {p,d,$^3$He}. In the following we analyze these initial density profiles for their eccentricities and their probability distributions, exploring the effects of the various sources of fluctuations (quark subdivision, gluon field fluctuations, multiplicity fluctuations) discussed in Sec.~\ref{sec2}. Since the results for x+Pb collisions at the LHC are qualitatively similar to those for x+Au collisions at top RHIC energy, with only minor quantitative differences, we focus in the following plots on the RHIC results.

%%%%%%%%%%%%%%%%%%%%%%%%%%%%%%%%%%%%%%%%%%%%%%%%%%
\subsection{Multiplicity fluctuations and centrality}
\label{sec3a}
%%%%%%%%%%%%%%%%%%%%%%%%%%%%%%%%%%%%%%%%%%%%%%%%%%

As already noted in the preceding section, the experimental procedure of ordering collision events by multiplicity and then cutting the set into ``centrality'' classes labelled by fractions of the total cross section (as illustrated in Fig.~\ref{F1}) produces in collisions involving small nuclei event classes with quite different characteristics than those we are used to from collisions between large nuclei. In collisions between large nuclei, such as the Pb+Pb collisions studied art the LHC, event multiplicity, the impact parameter $b$ and the number of participating (``wounded'') nucleons $N_\mathrm{part}$ are all tightly correlated, leading to uncertainties in $b$ and $N_\mathrm{part}$ of at most a few percent for fixed event multiplicity (see for example Fig.~7 in Ref.~\cite{Adam:2014qja}). This allows for a straightforward geometric interpretation of the measured multiplicity and justifies calling the resulting multiplicity bins ``centrality bins''. 

Fig.~\ref{F10} shows that correlations between multiplicity, $b$ and $N_\mathrm{part}$ become much weaker in collisions involving small nuclei.%
\footnote{%
It should be noted that multiplicity fluctuations and the use of realistic nucleon density profiles play a crucial role here, and in particular for p+p collisions Fig.~\ref{F10} would look completely different if a disklike collision criterion were used and multiplicity fluctuations were ignored, as is the case, for example, in the popular PHOBOS Glauber model \cite{Miller:2007ri,Loizides:2014vua}.
}
Due mostly to nucleon-nucleon multiplicity fluctuations, but also to the tails of the nucleon density profiles (whether modulated by quark-subdivision and/or gluon field fluctuations or not), cutting on fixed multiplicity produces event samples whose numbers of wounded nucleons and impact parameters fluctuate wildly. This basically invalidates a geometric interpretation of multiplicity as a measure of collision ``centrality'' in collisions involving small nuclei, in particular in p+p collisions. For this reason some experimental group have started replacing ``1 -- (collision centrality)'' by ``event activity'' for small collision systems, emphasizing the multiplicity-based nature of the experimental classification procedure.  

%
%%%%%%%%%%%%%%%%% Fig. 11 %%%%%%%%%%%%%%%%%%%%%%%%%%%%%%%
\begin{figure*}
	\includegraphics[width = 0.45\linewidth]{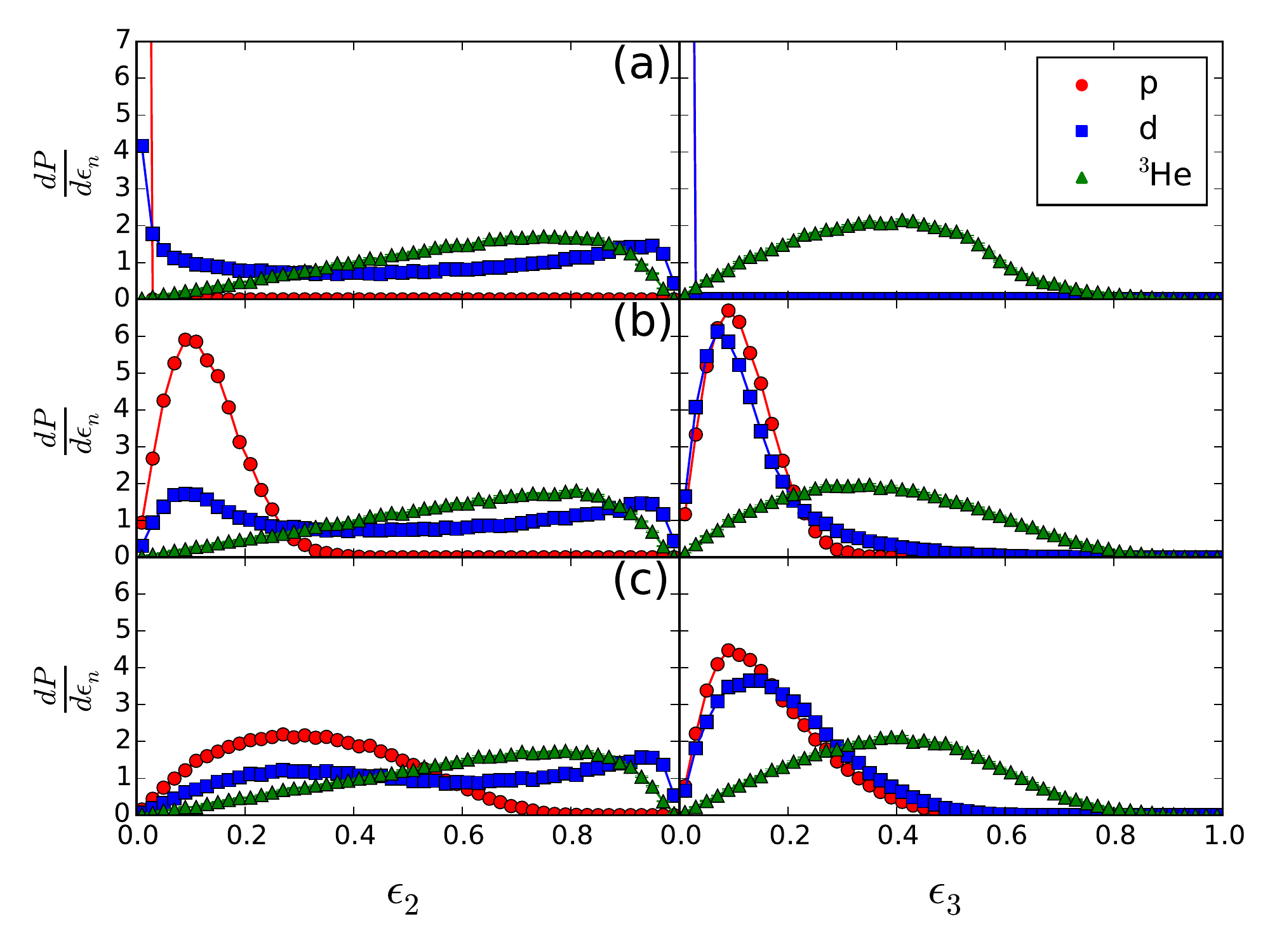}
	\includegraphics[width = 0.45\linewidth]{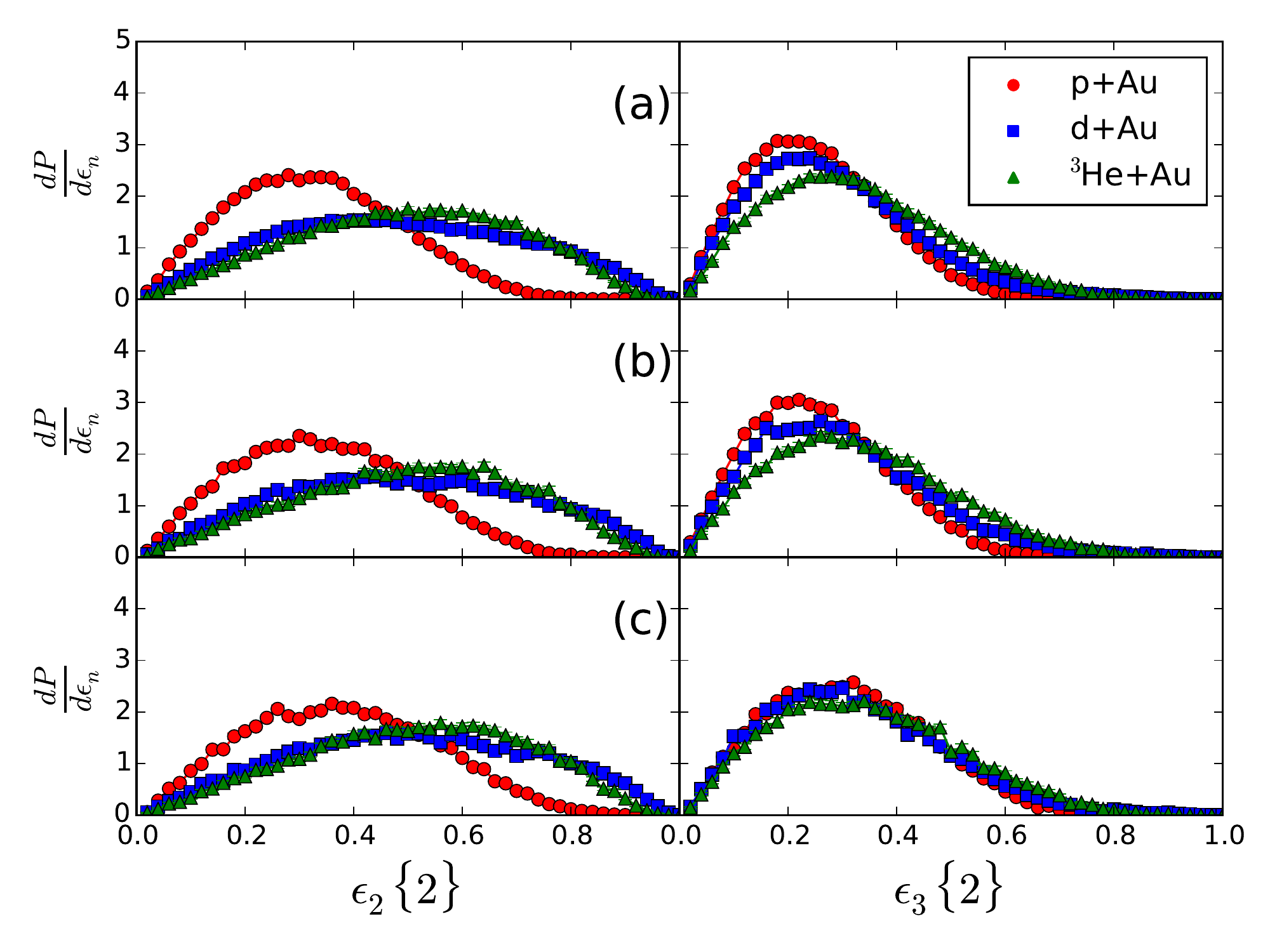}
	\caption{{\sl Left panel:} Probability distributions for the intrinsic eccentricities $\epsilon_2$ (left 
		column) and $\epsilon_3$ (right column) for protons (red circles), deuterons (blue squares), 
		and $^3$He nuclei (green triangles), calculated from the nuclear thickness functions for 
		smooth Gaussian nucleons (row (a)), Gaussian nucleons modulated by gluon field fluctuations 
		(row (b)) and quark-subdivided nucleons  modulated by gluon field fluctuations (row (c)). 
		{\sl Right panel:} Probability distributions for the same eccentricities calculated from the 
		deposited entropy density profile in central ($0-10\%$ centrality) p+Au (red circles), d+Au 
		(blue squares) and $^3$He+Au collisions at $\sqrt{s}\eq200\,A$\,GeV, including multiplicity 
		fluctuations, again using smooth Gaussian nucleons as well as Gaussian and 
		quark-subdivided nucleons modulated by gluon field fluctuations (rows (a)-(c), respectively).
		Without multiplicity and sub-nucleonic density fluctuations, we observed a significant 
		decrease of the mean ellipticity $\langle\epsilon_2\rangle$ for p+Au collisions and of the 
		mean triangularity $\langle\epsilon_3\rangle$ for d+Au collisions (not shown).
         \label{F11}
         }
\end{figure*}%
%%%%%%%%%%%%%%%%%%%%%%%%%%%%%%%%%%%%%%%%%%%%%%%%%%%
%

%%%%%%%%%%%%%%%%%%%%%%%%%%%%%%%%%%%%%%%%%%%%%%%%%%
\subsection{Intrinsic eccentricity distributions at impact}
\label{sec3b}
%%%%%%%%%%%%%%%%%%%%%%%%%%%%%%%%%%%%%%%%%%%%%%%%%%

Without multiplicity and sub-nucleonic density fluctuations, nucleons (by spherical symmetry) have no intrinsic eccentricities at all, deuterons (by reflection symmetry with respect to the plane spanned by the proton-neutron axis and the beam direction) have non-zero eccentricity coefficients only for even harmonic orders, and nonzero intrinsic triangularities are only possible for nuclei involving 3 or more nucleons (e.g. $^3$H or $^3$He). The entropy density profiles created in proton-proton collisions share the symmetries of the intrinsic deuteron thickness function and thus also have only even harmonic eccentricity coefficients.

As already mentioned in Sec.~\ref{sec2}, all of this changes when one allows for sub-nucleonic density fluctuations and pp multiplicity fluctuations. In Fig.~\ref{F11}a we show distributions of the elliptic and triangular eccentricities calculated from the nuclear thickness functions of samples of randomly oriented protons, deuterons and $^3$He nuclei (circles, squares and triangles). We call these (slight inaccurately) their ``intrinsic'' eccentricities. The figure explores the effects of gluon field fluctuations superimposed on Gaussian (row (b)) and quark-subdivided nucleons (row (c)) on these intrinsic ellipticity and triangularity distributions (left and right columns, respectively) and compares them to those obtained with smooth Gaussian nucleon density profiles in row (a).%
\footnote{%
	Note that pp multiplicity fluctuations do not enter here since they are only implemented in the 
	entropy deposition process after a collision has happened.
	}  
We see that subnucleonic fluctuations introduce non-zero ellipticities and triangularities in the proton, and that quark-subdivision increases the mean ellipticity of the proton but not its mean triangularity which remains (at the intrinsic level) much smaller than the mean triangularity of $^3$He nuclei. Quark subdivision shifts the ellipticity and triangularity distributions for deuterons to somewhat larger values, but doesn't change their mean values dramatically. The intrinsic triangularities of protons and deuterons have similar distributions and mean values, with and without quark subdivision -- gluon field fluctuations are sufficient to generate these.

The reader should note the large ``intrinsic'' ellipticities for $^3$He nuclei, with a mean value even larger than that of deuterons ($\langle\epsilon_2\rangle\eq0.60$ for $^3$He vs. 0.49 for d in row (a) which assumes smooth Gaussian nucleons). This is largely an effect of perspective: Although the three nucleons in the $^3$He nucleus are aligned in a plane, this plane is in general not oriented perpendicular to the beam axis. Even an equilateral triangle looks elongated when viewed at an angle. 

%%%%%%%%%%%%%%%%%%%%%%%%%%%%%%%%%%%%%%%%%%%%%%%%%%
\vspace*{-2mm}
\subsection{Eccentricity distributions directly after impact}
\label{sec3c}
\vspace*{-3mm}
%%%%%%%%%%%%%%%%%%%%%%%%%%%%%%%%%%%%%%%%%%%%%%%%%%

When a small nucleus hits a big target nucleus (such as Au), it generally wounds several target nucleons such that the total number of wounded nucleons (all of which contribute to the entropy density profile deposited in the collision) is significantly larger than the number of projectile nucleons. simple statistics suggests that this should reduce the effects of sub-nucleonic structure on the eccentricies of the deposited entropy profile relative to the ``intrinsic'' eccentricities of the projectile thickness functions. However, additional multiplicity fluctuations in the entropy deposition process counteract this tendency and are expected to increase the eccentricies of the deposited entropy profile relative to their ``intrinsic'' values. This is studied in Fig.~\ref{F11}b.

One sees that the ellipticities of the matter produced in p+Au collisions are generally much bigger than their intrinsic values in the projectile proton, eccept for the case of quark-subdivision. Almost no difference exists between the ellipticity distributions of the matter produced in d+Au and $^3$He+Au collisions. The large differences between the intrinsic triangularity distributions of p, d and $^3$He seen in Fig.~\ref{F11}a are almost completely washed out after the collision: The right column in Fig~\ref{F11}b shows that the triangularity distributions of the entropy profiles generated in central p+Au, d+Au and $^3$He+Au collisions are almost indistinguishable. In particular if one includes the effects of quark subdivision, protons generate matter with, on average, the same triangularity as $^3$He nuclei when colliding centrally with Au nuclei, contrary to initial expectations \cite{Nagle:2013lja} .

%%%%%%%%%%%%%%%%%%%%%%%%%%%%%%%%%%%%%%%%%%%%%%%%%%
\vspace*{-2mm}
\subsection{Centrality dependence of mean eccentricities and rms radii}
\label{sec3d}
\vspace*{-3mm}
%%%%%%%%%%%%%%%%%%%%%%%%%%%%%%%%%%%%%%%%%%%%%%%%%%

%
%%%%%%%%%%%%%%%%% Fig. 12 %%%%%%%%%%%%%%%%%%%%%%%%%%%%%%%
\begin{figure*}
	\includegraphics[width = 0.45\linewidth,height=5.6cm]{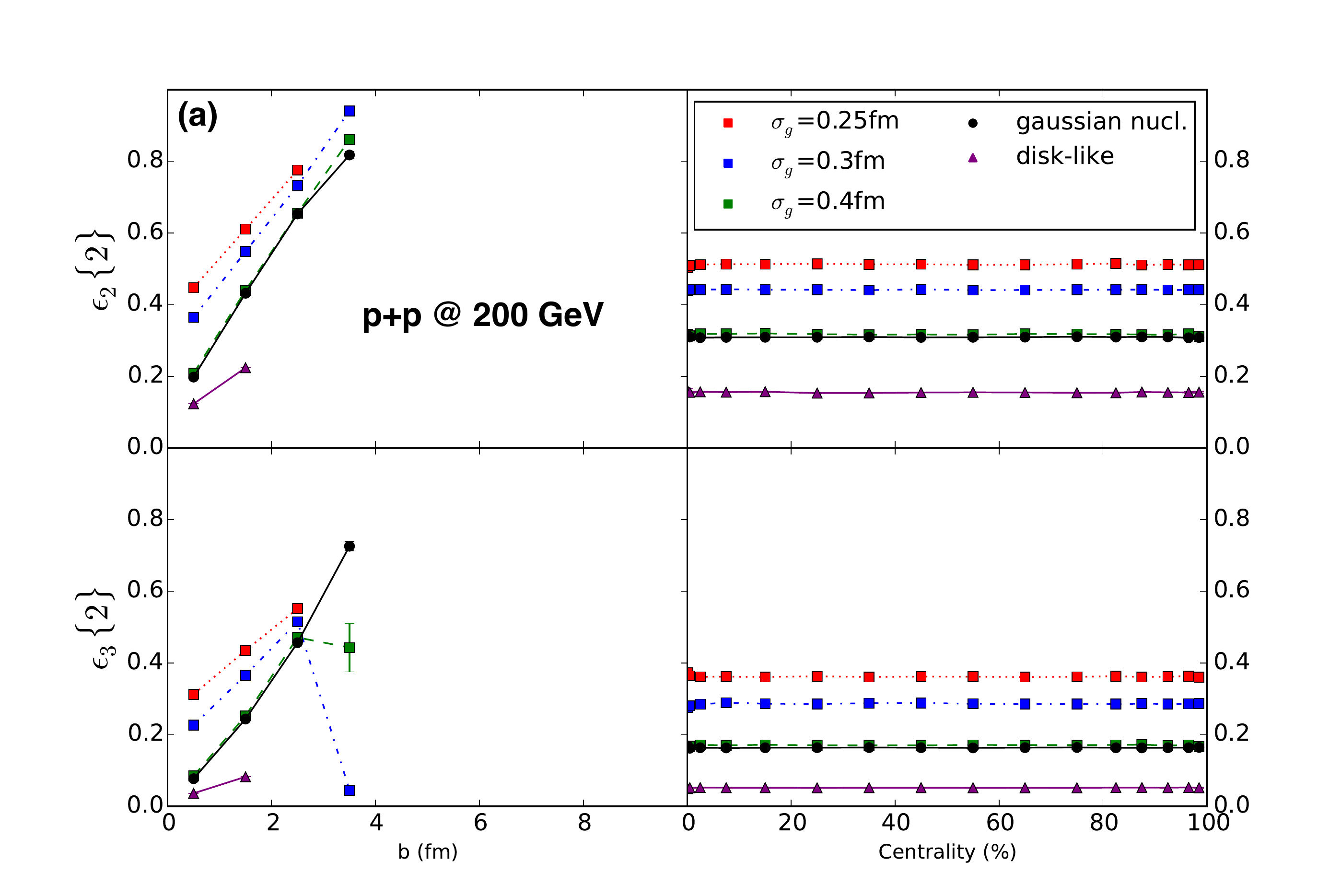}
	\includegraphics[width = 0.45\linewidth]{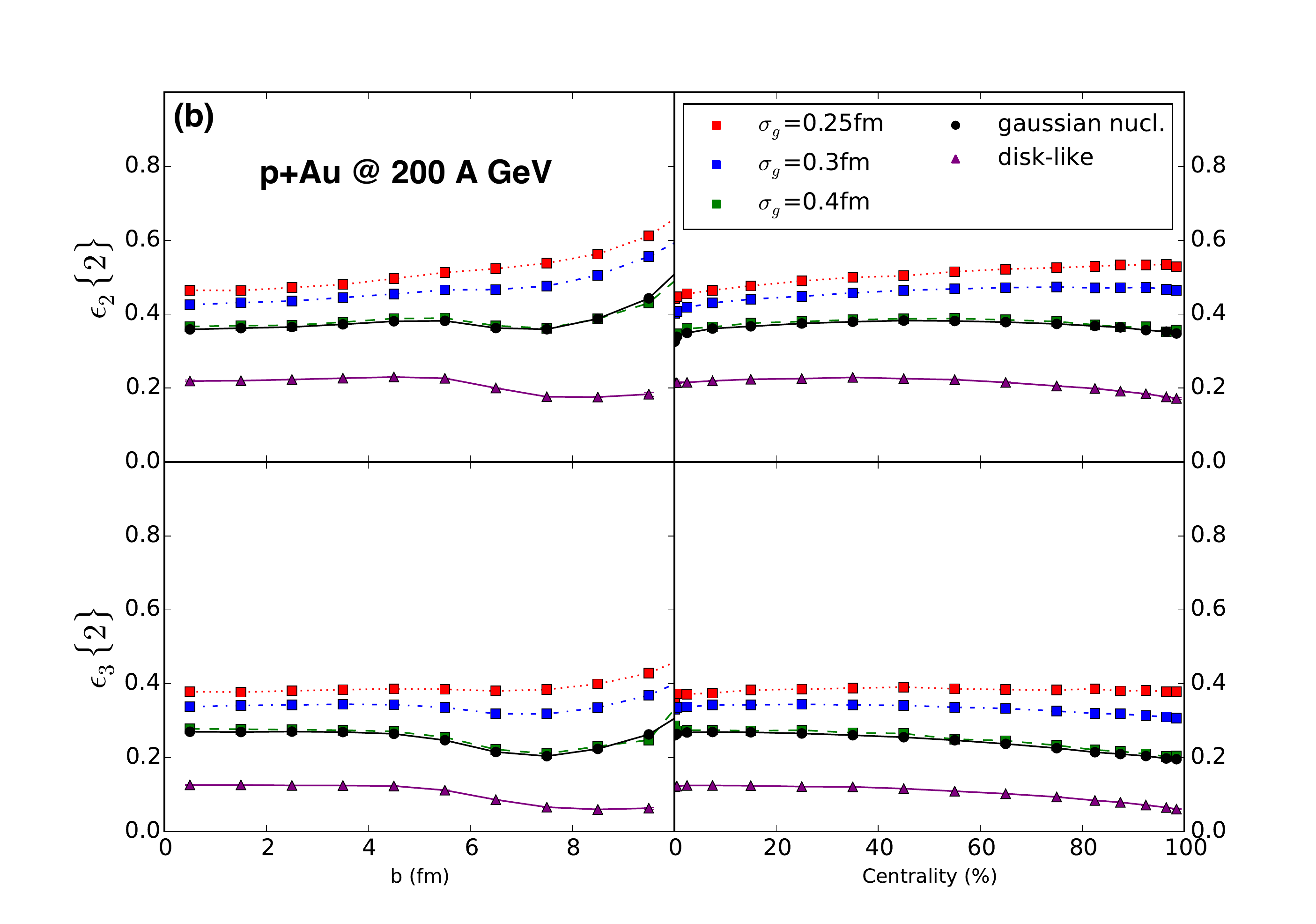}\\
	\includegraphics[width = 0.45\linewidth]{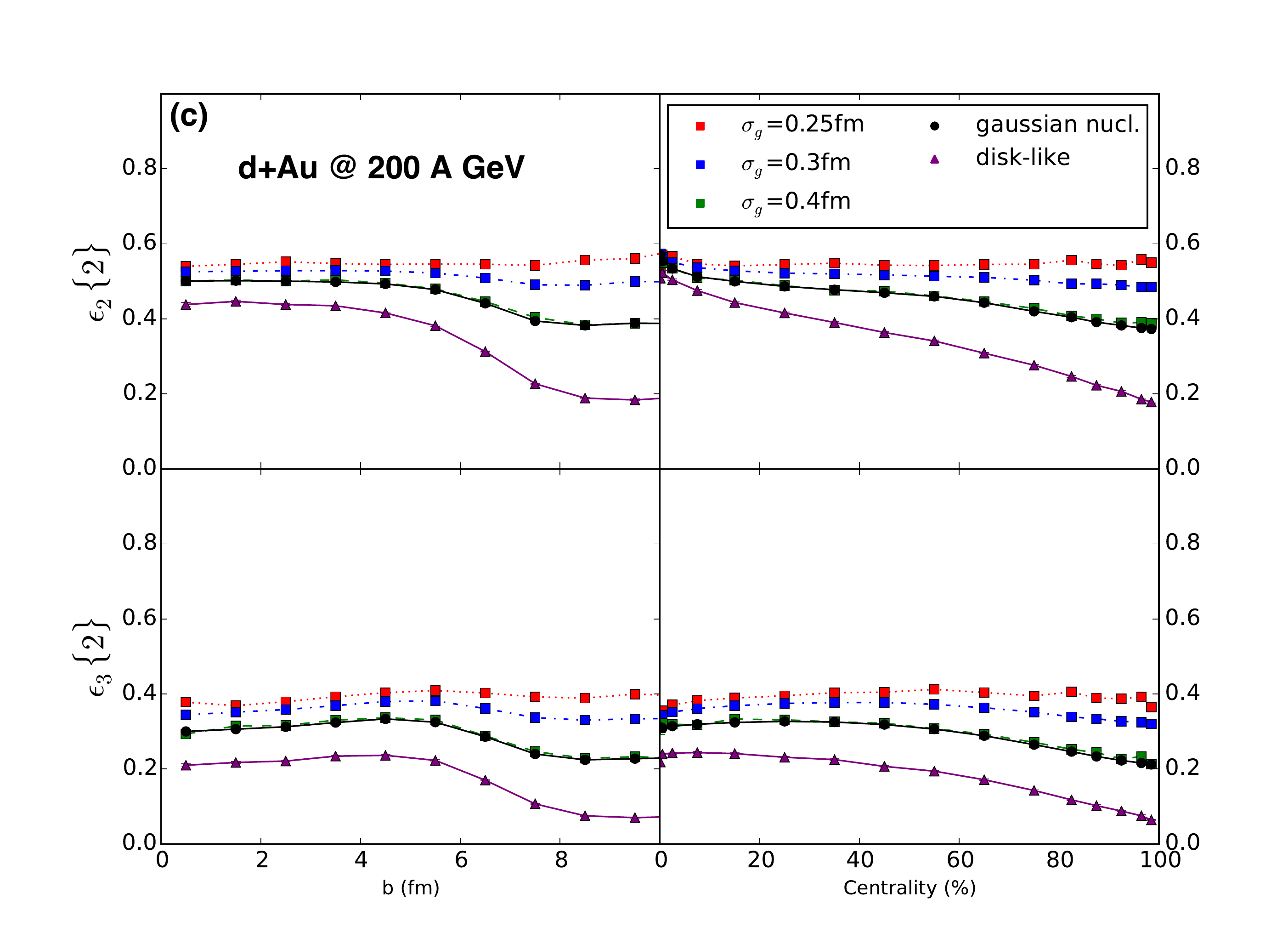}
	\includegraphics[width = 0.45\linewidth]{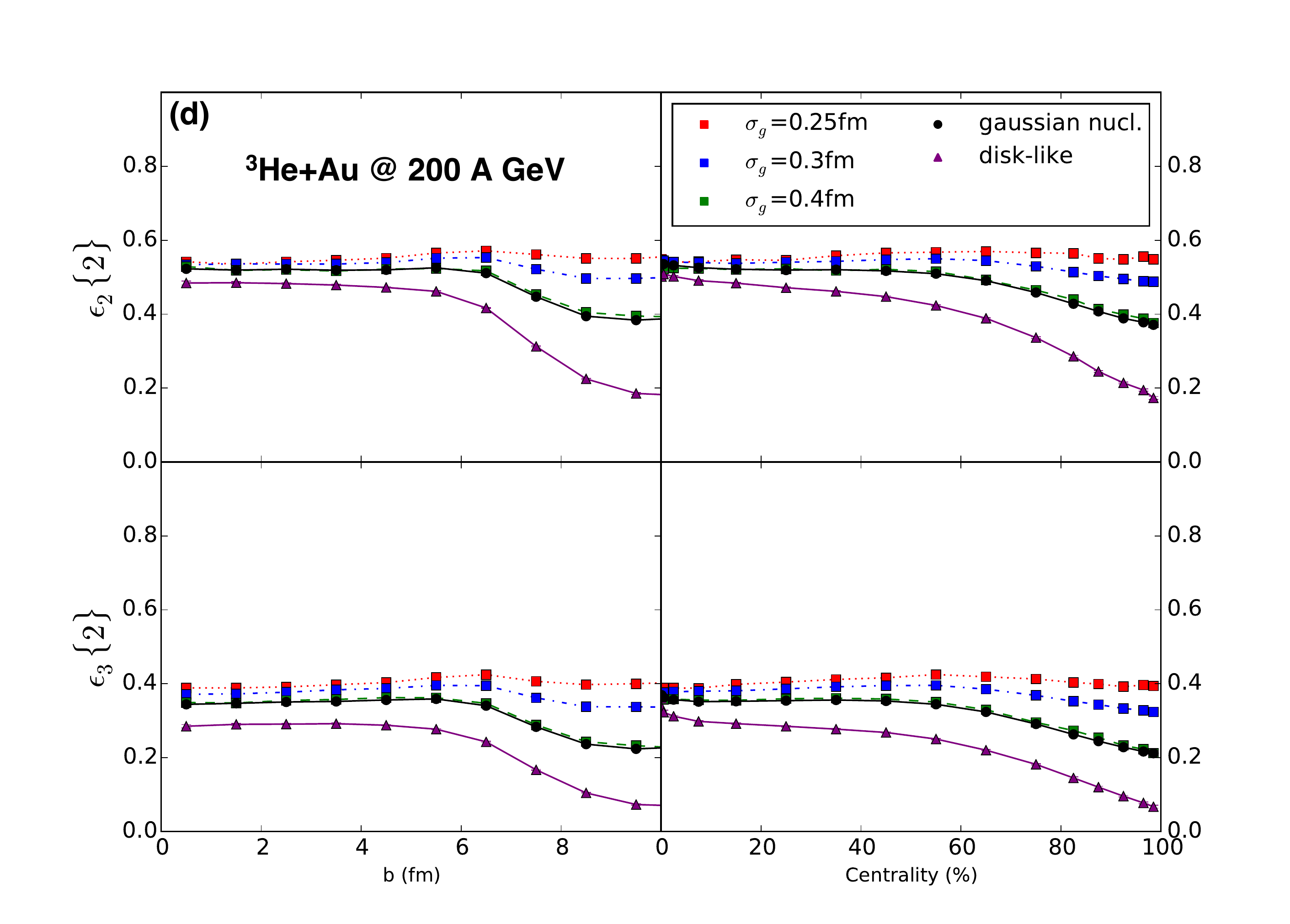}
	\caption{Centrality dependence of $\epsilon_{2}\{2\}$ (solid lines, shifted up by 0.2 for better
	 	visibility) and $\epsilon_{3}\{2\}$ (dashed lines) for p+p (a), p+Au (b), d+Au (c) and 
		$^3$He+Au (d) collisions at $\sqrt{s}\eq200\,A$\,GeV. In the left (right) panels impact 
		parameter (multiplicity) is used to characterize centrality. Purple triangles and black
		circles connected by solid lines show results from disk-like and Gaussian nucleon density
		profiles used in the collision detection algorithm. Red, blue and green squares connected by  
		dotted, dash-dotted and dashed lines, respectively, use quark-subdivided nucleon density
		profiles with widths $\sigma_g\eq0.25$, 0.3 and 0.4\,fm, respectively. All results include
		multiplicity fluctuations in the entropy deposition process. See text for discussion. 
         \label{F12}
         }
\end{figure*}%
%%%%%%%%%%%%%%%%%%%%%%%%%%%%%%%%%%%%%%%%%%%%%%%%%%%
%

In Figure~\ref{F12} we show the dependence of the elliptic and triangular eccentricities (upper and lower panels, respectively) for p+p and x+Au collisions at RHIC (x=p, d, $^3$He), as a function of impact parameter (left panels) and ``centrality'' as measured by multiplicity (right panels). Different curves correspond to standard disk-like collision detection, smooth Gaussian nucleons, and quark-subdivided nucleons with different widths of the valence quark gluon clouds, as described in the legend. Note the very different centrality dependences when centrality is defined in terms of impact parameter and multiplicity, respectively, especially for protons. This difference is caused by multiplicity fluctuations, as discussed in Sec.~\ref{sec3a}. Note that for Gaussian and quark-subdivided nucleons, the range of impact parameters in p+p collisions is limited by the reach of the Gaussian tails of the nucleon density distribution -- collisions at larger impact parameters are possible but so rare that our event sample contains too few events to allow for a meaningful calculation of $\epsilon_{2,3}\{2\}$. 

In collisions between large nuclei the initial triangularity and especially the eccentricity are known to exhibit significant dependence on the collision centrality. In contrast, we find for p+p and p+Au collisions almost no %
%%%%%%%%%%%%%%%%% Fig. 13 %%%%%%%%%%%%%%%%%%%%%%%%%%%%%%%
\begin{figure*}
	\includegraphics[width = 0.49\linewidth,height=5.55cm]{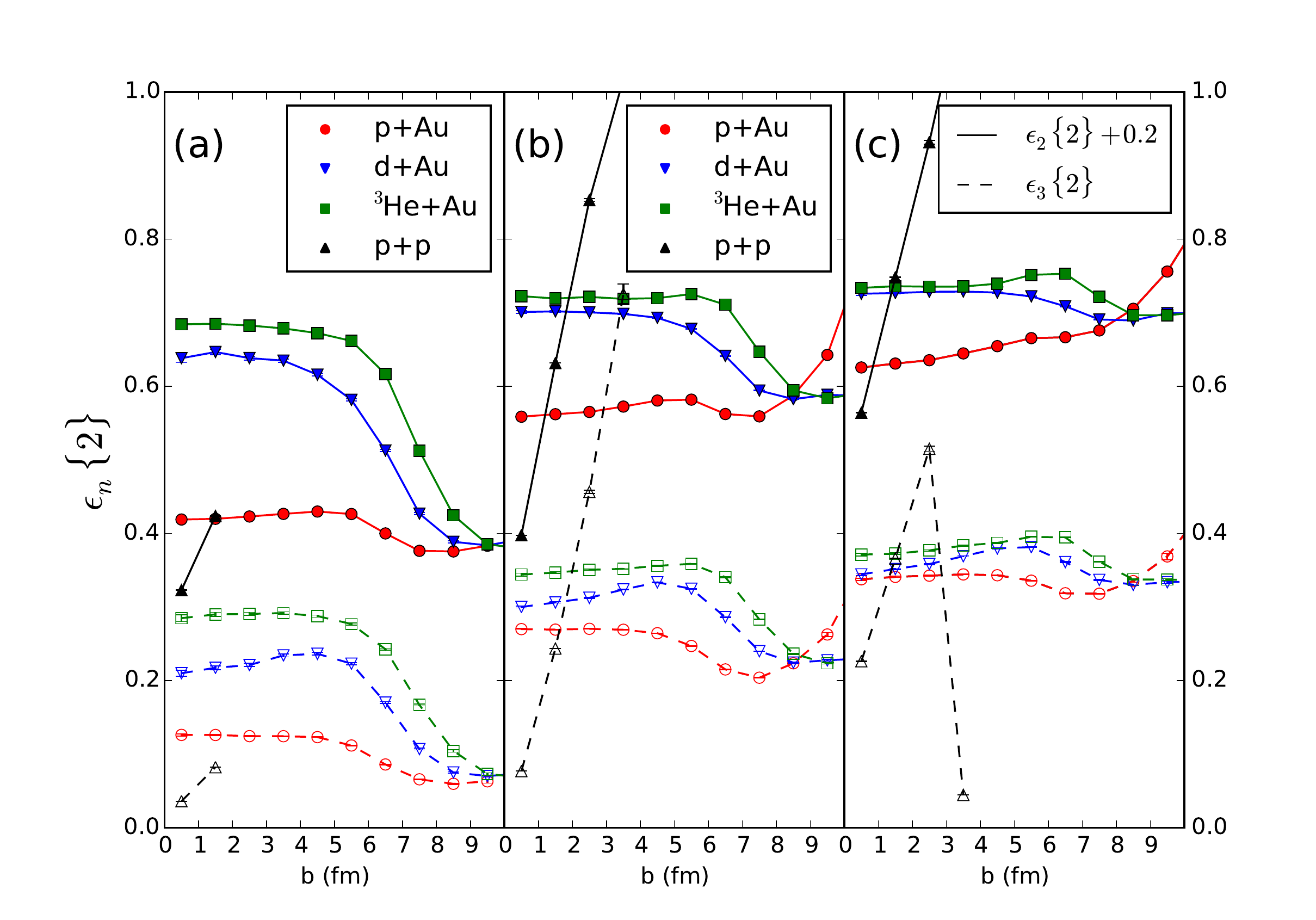}
	\includegraphics[width = 0.49\linewidth]{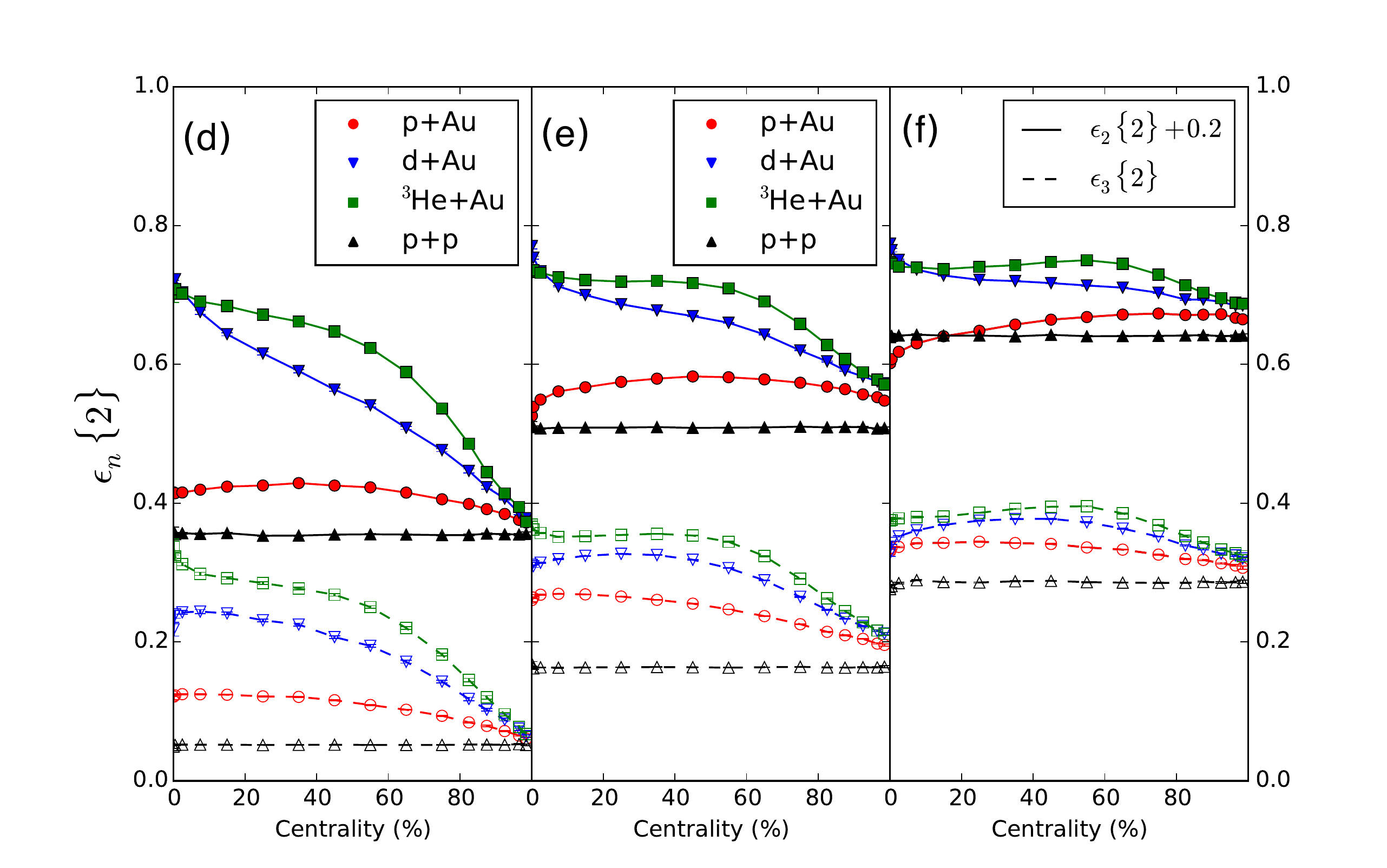}
	\caption{Similar to Fig.~\ref{F12}, but using different panels for different models for the collision 
	detection (disk-like (a,d), Gaussian (b,e) and quark-subdivided nucleons with $\sigma_g\eq0.3$\,fm
	(c,f)), and 	comparing in each panel different collision systems (black triangles: p+p; red circles: 
	p+Au; blue upside-down triangles: d+Au; green squares: $^3$He+Au). Panels a-c use impact 
	parameter, panels d-f use multiplicity to characterize collision centrality. See text for discussion.  
         \label{F13}
         }
\end{figure*}%
%%%%%%%%%%%%%%%%%%%%%%%%%%%%%%%%%%%%%%%%%%%%%%%%%%%
%
centrality dependence of $\epsilon_{2,3}\{2\}$ at all, and even for $^3$He+Au collisions the strong rise of $\epsilon_{2}\{2\}$ from central to mid-peripheral collisions known from Au+Au collisions is not yet visible; in fact, for disk-like collision detection and smooth Gaussian nucleons, both $\epsilon_2$ and $\epsilon_3$ decrease monotonically with increasing centrality. (This is more clearly visible in Fig.~\ref{F13} discussed below.) The most significant centrality dependence of $\epsilon_{2,3}\{2\}$ seen in the right panels of Figs.~\ref{F12}c,d is a strong decrease in low-multiplicity events (i.e. at large values for the centrality variable) for d+Au and $^3$He+Au collisions. Comparison with the left panels in the same figures and explicit study of the low-multiplicity events show that this decrease originates in geometrically highly peripheral (i.e. large impact parameter) collisions where, in the most extreme case, only a single valence quark with a strong upward fluctuation in its produced entropy dominates the deposited entropy profile.  

We note that sub-nucleonic shape fluctuations caused by quark subdivision reduce the centrality dependence of $\epsilon_{2,3}\{2\}$ in x+Au collision when the nucleus x is small. On the other hand, for all but the lowest-multiplicity events (i.e. for small to moderately large centrality values) the sensitivity of $\epsilon_{2,3}\{2\}$ on the width $\sigma_g$ of the valence quark clouds (and thus on the variance of the quark positions inside the nucleons) weakens significantly from p+p tp p+Au to $^3$He+Au collisions. In other words, the sub-nucleonic structure of protons and neutrons becomes increasingly irrelevant for the calculation of the initial-state eccentricities as the size of the colliding nuclei increases. Whereas for collisions involving small nuclei the initial eccentricities ``see'' the fluctuating internal spatial structure of each nucleon, for large collision systems essentially all that matters are the fluctuations on a nucleon size length scale caused by the fluctuations of the nucleon positions within the nucleus. While the latter observation has been made before \cite{...}, the strong sensitivity of the initial fireball geometry to details of the internal structure of the nucleon in small-on-small and small-on-large collisions is pointed out and systematically studied here for the first time.

In Figure~\ref{F13} we replot the results from Fig.~\ref{F12} in such away that the systematic change of the centrality dependence of $\epsilon_{2,3}$ with the collision system becomes more apparent. For both disk-like and smooth Gaussian nucleons (panels d and e) we see monotonic increases of both $\epsilon_2$ and $\epsilon_3$ with the size of the projectile x in central x+Au collisions, while in the most ``peripheral'' (lowest multiplicity) events their values approach the p+p values. It is no surprise that the increase of $\epsilon_2$ in central x+Au collisions is particularly strong for deuteron projectiles: In the most central collisions, the rms ellipticity in d+Au collisions even slightly exceeds the one for $^3$He+Au collisions. Comparison of panels d, e, and f (or, equivalently, a, b, and c) in Fig.~\ref{F13} shows that accounting for Gaussian tails in the nucleon density distribution and for quark substructure increase both $\epsilon_2$ and $\epsilon_3$ while at the same time reducing their centrality dependences.

%
%%%%%%%%%%%%%%%%%%%% Fig. 14 %%%%%%%%%%%%%%%%%%%%%%%
\begin{figure}[b]
	\includegraphics[width = \linewidth]{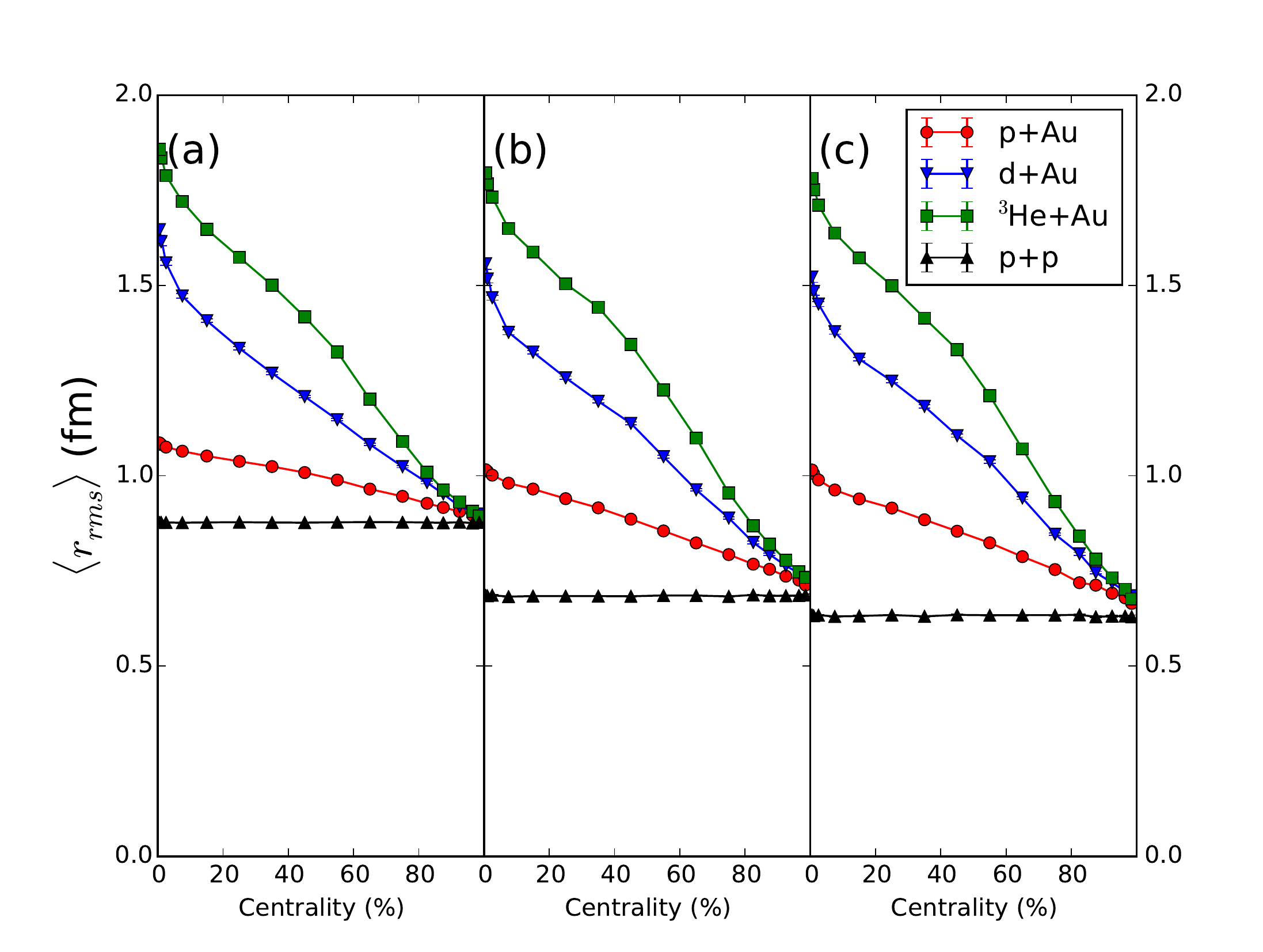}
	\caption{Similar to Fig.~\ref{F13}b, but for the mean rms radii of the initial entropy density
	   	distribution as functions of collision centrality, using different models for the collision 
		detection: disk-like (a), Gaussian (b), and quark-subdivided nucleons with 
		$\sigma_g\eq0.3$\,fm (c). In each panel different collision systems (black triangles: p+p; 
		red circles: p+Au; blue upside-down triangles: d+Au; green squares: $^3$He+Au) are 
		compared. See discussion in text. 
		\label{F14}
	}
\end{figure}%
%%%%%%%%%%%%%%%%%%%%%%%%%%%%%%%%%%%%%%%%%%%%%%%%
%

We note that Fig.~\ref{F13} exhibits significant qualitative differences compared to similar studies reported in \cite{Nagle:2013lja}. These differences demonstrate the importance of multiplicity and sub-nucleonic shape fluctuations in a complete description of the initial state of the fireballs created in p+p and x+Au/Pb collisions when x is small.   

%
%%%%%%%%%%%%%%%%% Fig. 15 %%%%%%%%%%%%%%%%%%%%%%%%%%%%%%%
\begin{figure*}
	\includegraphics[width = 0.9\linewidth]{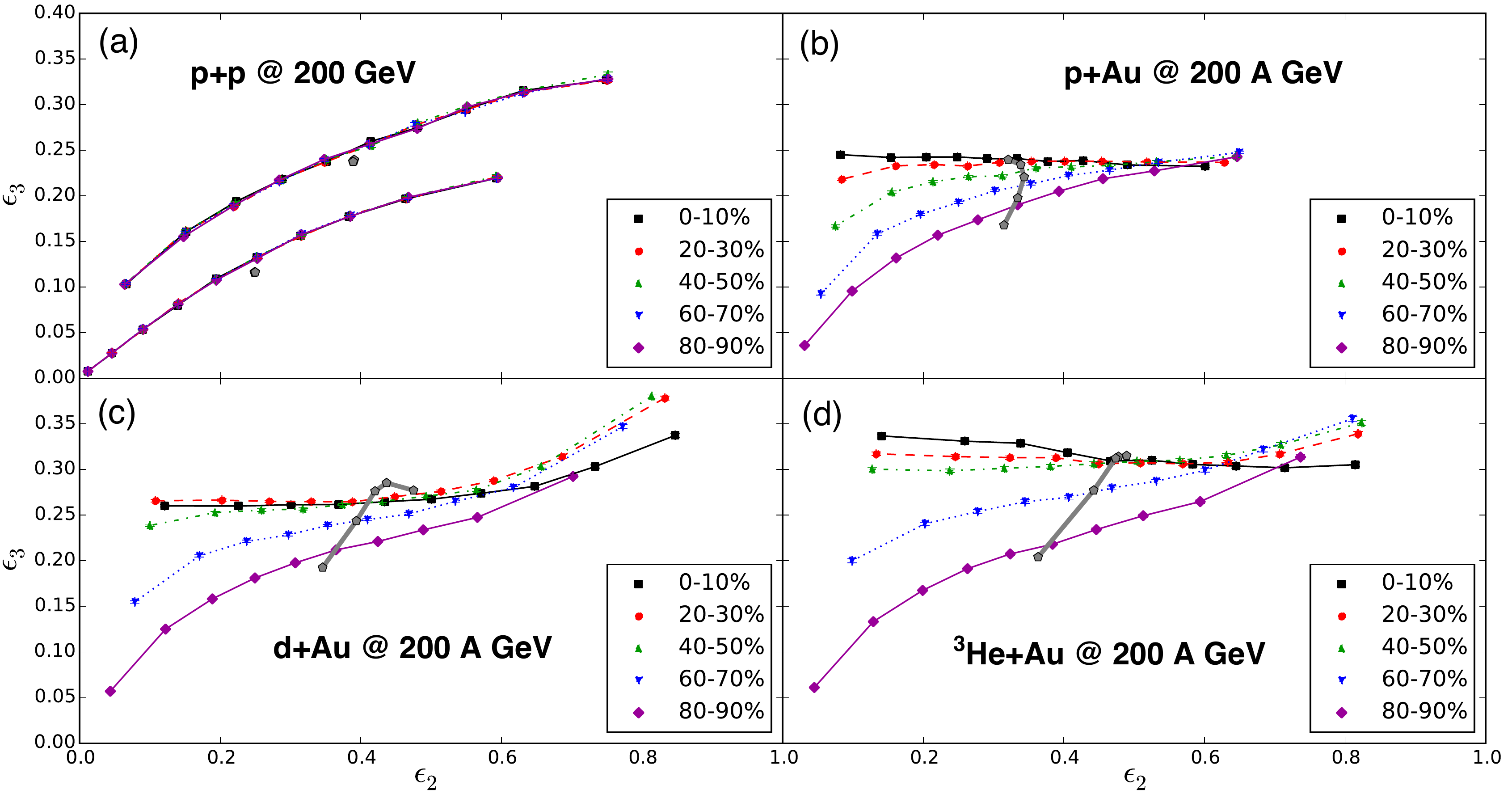}
	\caption{Eccentricity correlations between $\epsilon_3$ and $\epsilon_2$ in p+p (a), 
	p+Au (b), d+Au (c) and $^3$He+Au collisions (d) at $\sqrt{s}\eq200\,A$\,GeV. The calculations 
	assume quark-subdivided nucleons with gluon clouds of width $\sigma_g\eq0.3$\,fm and include
	multiplicity fluctuations in the entropy deposited by each struck valence quark. The gray line in
	the center of each panel (which degenerates to a point in panel (a) due to the centrality 
	independence of both $\epsilon_2$ and $\epsilon_3$ in p+p collisions) connect the mean 
	values of $\epsilon_3$ vs. $\epsilon_2$ for the centrality bins shown in the legend. The colored
	points subdivide the events in each centrality bin into bins with different $\epsilon_2$ values and 
	plot the mean triangularity $\langle\epsilon_3\rangle$ vs. the mean ellipticity 
	$\langle\epsilon_2\rangle$ in each such bin. See text for discussion.	
         \label{F15}
         }
\end{figure*}%
%%%%%%%%%%%%%%%%%%%%%%%%%%%%%%%%%%%%%%%%%%%%%%%%%%%
%

We close this subsection by showing in Fig.~\ref{F14} the centrality dependence of the mean rms radii as functions of collision centrality for p+p, p+Au, d+Au, and $^3$He+Au collisions at $\sqrt{s}\eq200\,A$\,GeV, for the same three collision detection models studied in Fig.~\ref{F13}. While for p+p collisions the
rms radii are approximately independent of collision centrality as measured by multiplicity (as already shown in Fig.~\ref{F6}), collisions of protons or small nuclei on Au targets produce much larger initial source sizes in high-multiplicity than in low-multiplicity events. At the left end of the centrality distribution (i.e. for the highest multiplicity events) the choice of collision criterium and quark subdivision have negligible effect on the initial fireball radius in x+Au collisions even though their effect is strong in p+p collisions. At the right end on the centrality distribution (i.e. for the lowest produced multiplicities) the sources produced in x+Au collisions have the same size as those produced (at any centrality) in p+p collisions. The smaller initial sources in ``peripheral'' (i.e. low-multiplicity) x+Au collisions feature larger initial pressure gradients, resulting in stronger hydrodynamic acceleration, but smaller initial entropy content, resulting in shorter lifetimes and earlier freeze-out. In future studies of the hydrodynamic evolution of the initial configurations analyzed in the present work, it will be interesting to explore the consequences of the interplay between these counteracting effects on the final mean $p_T$ values of hadrons emitted from central vs. peripheral x+Au collisions. 

%%%%%%%%%%%%%%%%%%%%%%%%%%%%%%%%%%%%%%%%%%%%%%%%%%
\vspace*{-2mm}
\subsection{Eccentricity correlations}
\label{sec3e}
\vspace*{-3mm}
%%%%%%%%%%%%%%%%%%%%%%%%%%%%%%%%%%%%%%%%%%%%%%%%%%

In Figure~\ref{F15} we present an analysis of eccentricity correlations, specifically of correlations between the magnitudes of the ellipticity $\epsilon_2$ and triangularity $\epsilon_3$, inspired by a similar analysis for the corresponding flow coefficients first performed on experimental data from Pb+Pb collisions at the LHC by the ATLAS collaboration \cite{Aad:2015lwa} and recently repeated on the results from hydrodynamic model calculation in \cite{Qian_new}. Due to the well-established linearity of the elliptic (triangular) flow response to $\epsilon_2$ ($\epsilon_3$), we expect the $v_3$-$v_2$ correlations to look similar to Fig.~\ref{F15} once the initial entropy density profiles generated in this work will have been evolved hydrodynamically.

The grey band in the center of each panel in Fig.~\ref{F15} connects the pairs $(\langle\epsilon_2\rangle,\langle\epsilon_3\rangle)$ for each of the centrality bins shown in the legend. (In panel (a) this band degenerates to a point, due to the centrality independence of both $\epsilon_2$ and $\epsilon_3$ in p+p collisions.) For the colored points we ordered the events in each centrality class by ellipticity $\epsilon_2$ and subdivided them into 10 equally occupied ellipticity bins. What is plotted in Fig.~\ref{F15} is the mean triangularity $\langle\epsilon_3\rangle$ of the events in each of these bins vs. their mean ellipticity $\langle\epsilon_2\rangle$, connected by lines for each centrality class and separated by color, symbols and line style for different centrality classes. 

In p+p collisions we observe a simple pattern of event triangularities that increase monotonically with event ellipticity, at a rate that is completely independent of collision centrality. This is caused by the large effects from multiplicity fluctuations which largely wash out all geometric differences between different centrality classes.

For x+Au collisions, where x stands for p (Fig.~\ref{F15}b), d (Fig.~\ref{F15}c) or $^3$He (Fig.~\ref{F15}d), we observe triangularities that are essentially uncorrelated with $\epsilon_2$ in ``central'' collisions but become more and more strongly positively correlated with $\epsilon_2$ as the centrality increases (i.e. the multiplicity decreases). As the collision centrality approaches 100\% the correlation between $\epsilon_2$ and $\epsilon_3$ becomes similar to the one seen in p+p collisions. 

We point out that the characteristic patterns seen in Fig.~\ref{F15} for p+p and small-on-large collisions are opposite to those seen in large-on-large collisions such as Pb+Pb \cite{Aad:2015lwa,Qian_new}: While in all cases $\epsilon_3$ (or $v_3$) is uncorrelated with $\epsilon_2$ (or $v_2$) in central collisions, $\epsilon_3$ ($v_3$) develops at non-zero centrality a {\em negative (anti-) correlation} with $\epsilon_2$ ($v_2$) in non-central Pb+Pb collisions, instead of the {\em positive correlation} observed here for p+p and small-on-large collisions. The anti-correlation between $\epsilon_3$ and $\epsilon_2$ observed in non-central Pb+Pb collisions has been attributed in \cite{Aad:2015lwa,Qian_new} to geometric deformation effects of the nuclear overlap region. Since centrality in p+p and small-on-large collisions cannot be interpreted geometrically, such a geometric anti-correlation is not visible in Fig.~\ref{F15}.

%%%%%%%%%%%%%%%%%%%%%%%%%%%%%%%%%%%%%%%%%%%%%%%%%%
\section{Summary and conclusions}
\label{sec4}
\vspace*{-2mm}
%%%%%%%%%%%%%%%%%%%%%%%%%%%%%%%%%%%%%%%%%%%%%%%%%%

In this work we studied the initial elliptic and triangular eccentricity coefficients of the mid-rapidity matter produced in high-energy collisions involving small projectile and/or target nuclei. We showed that, contrary to collisions between large nuclei, in such small-on-small or small-on-large collisions sub-nucleonic density fluctuations play a crucial role for the size and shape of the initially produced matter, and that in particular accounting for multiplicity fluctuations completely changes the centrality dependences of the initial ellipticities and triangularities (where centrality is defined in terms of the multiplicity of produced particles, as done in experiment). Our results suggest that a quantitative understanding of recent experimental measurements of anisotropic flow in high-multiplicity p+p and in p+Au, d+Au, $^3$He+Au and p+Pb collisions at RHIC and LHC requires a careful and systematic study of the initial state of the hot matter created in these collisions, and of its effect on the dynamical evolution of that matter.

The results presented here focus entirely on the initial state of the produced matter. For comparison with experiment these initial conditions must be propagated dynamically to the experimentally observed final state, using a dynamical evolution model. Results from ongoing hydrodynamic simulations using the {\tt iEBE-VISHNU} code package \cite{Shen:2014vra} will be reported separately.

\acknowledgments{%
\vspace*{-2mm}
We acknowledge fruitful and stimulating discussion with Piotr Bo\.zek, Wojciech Broniowski, Heikki M\"antysaari, and Jamie Nagle, and thank Joe Carlson and Joel Lynn for providing us with 14,000 sampled nucleon configurations for $^3$He nuclei using state-of-the-art 3-nucleon wave functions. We thank Brian Baker for valuable contributions made during the initial stages of this project while supported by an undergraduate summer research scholarship from the OSU Department of Physics. KW gratefully acknowledges support for two summers through undergraduate research scholarships from the Department of Physics at The Ohio State University. This work was supported by the U.S. Department of Energy, Office of Science, Office of Nuclear Physics under Award No. \rm{DE-SC0004286}.}

%%%%%%%%%%%%%%% References %%%%%%%%%%%%%%%%%%%%%%%%%%%

\end{document}